\documentclass[journal]{IEEEtran}

\usepackage{amsmath}
\usepackage{amssymb}
\usepackage{xcolor}
\usepackage{soul}
\usepackage{algpseudocode}
\usepackage{algorithm}
\usepackage{setspace}
\usepackage{enumitem,kantlipsum}

\usepackage[noadjust]{cite}
\usepackage [english]{babel}
\usepackage [autostyle, english = american]{csquotes}
\usepackage{graphicx}
\usepackage{subcaption}
\usepackage{epsf}
\usepackage{epstopdf}
\usepackage{float}
\usepackage{stfloats}
\MakeOuterQuote{"}

\begin{document}
\title{Joint Uplink-Downlink Resource Allocation for Multi-User IRS-Assisted Systems}

\author{Mahmoud Saad Abouamer,~\IEEEmembership{Student Member,~IEEE}, Patrick Mitran,~\IEEEmembership{Senior Member,~IEEE}
        
\thanks{Mahmoud Abouamer and Patrick Mitran are  with the Department
of Electrical and Computer Engineering, University of Waterloo, Waterloo,
ON, Canada. e-mail: $\{$mabouamer and pmitran$\}$@uwaterloo.ca.} 
 \thanks{This work was supported in part by the Natural Sciences and Engineering Research Council of Canada (NSERC) and Cisco.}  

}
\maketitle

\begin{abstract}
\textcolor{black}{We investigate the joint uplink-downlink configuration of an intelligent reflecting surface (IRS) for multi-user frequency-division-duplexing (FDD) and time-division-duplexing (TDD) systems. This is motivated in FDD since uplink and downlink transmissions occur simultaneously and hence an IRS must be jointly configured for both transmissions. In TDD, while a {joint design} is not strictly necessary, it can significantly reduce feedback overhead, power consumption, and {configuration periods} associated with updating the IRS. To compute the trade-off between uplink and downlink rates achieved by a {joint design}, a weighted-sum problem is formulated and optimized using a developed block-coordinate descent algorithm. \textcolor{black}{The resulting uplink-downlink trade-off regions are investigated by numerical simulation to gain insights into different scenarios.} In all FDD scenarios and some TDD scenarios, the jointly optimized design significantly outperforms the fixed-uplink (fixed-downlink) heuristic of using the IRS configuration optimized for uplink (downlink) to assist downlink (uplink) transmissions. Moreover, the {joint design} substantially bridges the gap to the {individual design} upper bound of allowing different IRS configurations in uplink and downlink. \textcolor{black}{Otherwise, in the remaining TDD scenarios, the fixed-uplink and fixed-downlink designs nearly achieve the {individual design} performance and substantially reduce overhead and/or complexity compared to the {optimized joint design} and {individual design}.}}  
\end{abstract}
\vspace{-0.25cm}
\section{Introduction}
\par
The relentless demand for increased wireless connectivity, data rates, and quality-of-service (QoS) requirements necessitates improvements in spectrum efficiency, power efficiency, and reliability. \textcolor{black}{Current technologies focus on designing the transmitter or the receiver, and the configuration of the wireless channel itself remains largely beyond the reach of the designer.} Intelligent reflecting surfaces (IRS) can fill this gap by enhancing the performance of a communication system by tuning the wireless propagation channel. Specifically, an IRS encompasses a large array of configurable reflecting elements that can collectively reshape the incident signal in terms of its phase and amplitude \cite{IRS_SDN,IRS-Magazine,IRSvsRelay,opport}.

\par 
\textcolor{black}{{A key design aspect of IRS is the need to configure the reflecting elements to achieve a desired objective.} The literature on IRS spans a wide variety of system design objectives including maximizing the weighted sum-rate (WSR) \cite{IRS-MO}, minimizing the total power subject to SINR constraints \cite{IRS_power_min,power-control-IRS}, physical-layer security \cite{Security-2}, and improving capacity for indoor and outdoor settings \cite{IRS-capacity,Indoor-capacity}. The rate optimization of IRS-assisted MIMO systems and MIMO BC channel were discussed in \cite{ARO} and \cite{MIMO-BC}, respectively. Furthermore, the interplay between IRS designs and other emerging wireless paradigms such as non-orthogonal multiple access \cite{NOMA}, full duplex (FD) communication \cite{Fd_IRS_2} and simultaneous wireless information and power transfer \cite{IRS_SWIPT} is discussed in several recent papers. {Due to the unit-modulus constraint imposed on the IRS phase shifts and the coupling of decision variables, the optimization problems associated with {joint beamforming and IRS configuration} are non-convex. Consequently, Riemannian conjugate gradient (RCG) and the weighted minimum mean square error  (WMMSE)  algorithm have been adopted in early IRS literature (e.g. \cite{early_IRS_MO,early_IRS_WMMSE}) as well as more recent works (\cite{MO_recent_1,MO_recent_2,MO_recent_3,position_IRS}) to jointly optimize the IRS phase shifts and beamforming vectors.}} 
\par 
\textcolor{black}{In this paper, we consider the practical problem of jointly optimizing the uplink WSR and the downlink WSR for multi-user (MU) multiple-input single-output (MISO) FDD and TDD systems.  We adopt a joint uplink-downlink IRS design where the same IRS configuration is used to assist both uplink and downlink transmissions in FDD and TDD systems {and where the time/frequency resources dedicated to uplink/downlink can be unequal.} \textcolor{black}{By leveraging RCG, WMMSE, and fractional programming (FP), trade-off regions between uplink and downlink rates achieved by a jointly optimized design are computed and compared to the regions generated by the fixed-uplink/fixed-downlink heuristic designs and the individual design upper bound of allowing different IRS configurations for uplink and downlink.}} For FDD systems, using the same IRS configuration is critical in order to support simultaneous uplink (UL) and downlink (DL) transmissions as they occur over the same time resources \cite{FDD-IRS}. On the other hand, while it is feasible for TDD systems to support different IRS configurations for uplink and downlink transmissions, a joint IRS design can trade separately optimized IRS phase shifts for reduced signalling overhead, power consumption and delay associated with updating the IRS. \textcolor{black}{ This trade-off may be particularly appealing when the loss in spectral efficiency due to a joint design compared to the individual design is marginal, as will be demonstrated for many TDD scenarios considered in Section \ref{sec 5}.} 
\par
 \textcolor{black}{ Feedback overhead and delay associated with updating an IRS are among the major challenges that face IRS implementations in practical scenarios \cite{IRS-synch, two-timescale,partial_CSI,overhead_basic,Overhead_aware,position_IRS,Myth}. The feedback duration may significantly reduce the time available for data transmission and hence hinders the performance enhancements brought by deploying an IRS \cite{Overhead_aware,position_IRS}. {In addition, whether an IRS can be controlled in real time, especially in mobile environments, remains a critical issue \cite{Myth}. Consequently, depending on the technology,  quiet guard periods while updating the IRS may be required to avoid the time-varying behavior of the channel due to a changing IRS configuration.}\footnote{ \textcolor{black}{For instance, micro-electro-mechanical systems (MEMS) {is one technology} envisioned for IRS implementation as they are cost-effective, have high quality factors and consume relatively low power \cite{MEMS_comm_1,MEMS_material_1}. However, one of the main drawbacks of MEMS is their slow configuration speed \cite{MEMS_material_1,MEMS_material_2}.}} {Subsequently, there is a  time where an IRS may not be used for communication due to feedback and quiet guard periods which we refer to jointly as the \emph{configuration period}.} \textcolor{black}{In TDD,  compared with separate individual designs for uplink and downlink, a joint design reaps the benefits of halving the feedback overhead as well as \textcolor{black}{reducing} the configuration periods. Furthermore, a joint IRS design can also \textcolor{black}{reduce} the power consumption of the IRS by \textcolor{black}{reducing} the need to switch the IRS configuration.} This follows as most of the power consumed by an IRS is used in controlling and configuring the IRS \cite{Myth}. } 
 \par
\begin{figure}[t]
        \includegraphics[width=\linewidth]{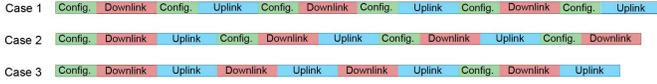}
    \caption{{Three TDD cases showing {possible} allocation of time slots in Individual design (case 1) and joint design (cases 2 and 3). \textcolor{black}{Here, the update overhead and duration are reduced in cases 2 and 3 compared to case 1.} }}
    \label{fig:TDD time slots}
\end{figure}

 \textcolor{black}{Fig.~\ref{fig:TDD time slots} shows three possible update schedules for TDD operation: in case 1 the conventional approach of configuring the IRS before every uplink/downlink transmission; in case 2 the joint uplink/downlink approach where the \textcolor{black}{IRS is only configured before each downlink transmission;} \textcolor{black}{and if the channel coherence time is long {enough}, then a joint design may require no update over several uplink/downlink intervals as shown in case 3 of Fig.~\ref{fig:TDD time slots}, thus further reducing configuration periods and power consumption.}}
\par
\textcolor{black}{Joint uplink-downlink design for IRS was discussed in \cite{FDD-IRS} in the context of an FDD system in which a single-antenna  BS is used to serve a single-antenna user and equal bandwidth was allocated to uplink and downlink transmissions. {{The use of an IRS configured for uplink to assist  downlink transmission was suggested for the (interference-free) case when a single-antenna BS serves a single-antenna user under TDD scenario \cite{reciprocity-2}}}.  Moreover, joint design also arises in full-duplex systems such as \cite{Fd_IRS_2} and \cite{Fd_IRS_1} where the IRS is used, in part, to improve the performance of the system by suppressing self-interference.} 
\par 
\textcolor{black}{In this work, a joint uplink-downlink design is proposed to support concurrent transmissions in practical multi-user IRS-assisted FDD systems as well as reduce overhead and configuration periods in practical multi-user IRS-assisted TDD systems. Moreover, two heuristic designs and two slicing benchmarks are devised in order to assess the performance of the proposed joint design under different scenarios. Subsequently, state-of-the-art algorithms are adopted to efficiently optimize the proposed joint design and two heuristic designs and investigate trade-off regions between achievable downlink and uplink rates in IRS systems. In particular, the uplink-downlink trade-off regions are investigated by numerical simulation for different user priorities, beamforming strategies and FDD/TDD system parameters. Furthermore, we analyze the complexity associated with the proposed joint design and heuristic designs. 
This analysis is then used along with the computed trade-off regions to illustrate the trade-offs associated with different IRS designs in terms spectral efficiency performance, configuration periods, overhead and computational complexity.} We summarize the contributions and key findings of this paper as follows: 
\begin{itemize}[leftmargin=*]
\item \textcolor{black}{We investigate joint uplink-downlink IRS design in multi-user multiple-input-single-output FDD and TDD systems. By leveraging RCG, WMMSE and FP, an optimized joint IRS design trade-off region between UL and DL is obtained when the same IRS configuration is used.}
\item \textcolor{black}{To evaluate the performance of the optimized joint UL-DL design, we compare it to optimizing the IRS for UL and DL transmissions individually and then time-sharing these two IRS configurations, referred to as \emph{Individual design}. \textcolor{black}{Moreover, we propose two heuristic designs called \emph{fixed-downlink} and \emph{fixed-uplink}. In addition, two slicing benchmarks called \emph{slicing-without-interference} and \emph{slicing-with-interference} are also introduced. }} 
\item \textcolor{black}{ \textcolor{black}{The impact of different user-weighting strategies \textcolor{black}{(equal, proportional-fair, and independent weights)} on the uplink-downlink trade-off regions  associated with the jointly optimized design and two fixed designs is analyzed by numerical simulation.} It is found that the improvement due to joint design compared with the fixed-downlink/fixed-uplink designs becomes more significant as{ the user-weights become less uniform.}} 
\item \textcolor{black}{\textcolor{black}{The impacts of both the ratio of IRS elements to active antennas at the BS and beamforming strategy (i.e., {WMMSE}, {{zero-forcing (ZF)}}) on the trade-off regions are also investigated. As the ratio of IRS to BS elements is increased,} the performance improvement due to the joint IRS design over fixed-uplink/fixed-downlink designs becomes more pronounced. Moreover, it is found that the improvement in performance due to joint design compared with the fixed  uplink/downlink designs is similar under the two considered beamforming strategies.}
\item \textcolor{black}{ The uplink-downlink trade-off analysis demonstrates that the region achieved by an optimized joint design strictly contains the regions due to slicing-without-interference and slicing-with-interference. This shows not only that a joint IRS design outperforms the slicing benchmarks but also demonstrates that interference alone is not responsible for the inferior performance achieved by a sliced IRS.}
\item \textcolor{black}{The complexity associated with the optimized joint IRS design and {fixed designs} is analyzed. This analysis is then used along with the computed trade-off regions to gain insights into scenarios where a marginal loss in spectral efficiency can be traded off for reductions in configuration periods, overhead and complexity.}
\item \textcolor{black}{In FDD, the proposed jointly optimized IRS design performs substantially better than {the fixed-uplink and fixed-downlink heuristic designs} {and diminishes the gap to the individual design}. \textcolor{black}{In TDD, provided that the user weights are independent, the joint design substantially improves the spectral efficiency performance compared with the fixed-uplink and fixed-downlink heuristic designs.} \textcolor{black}{Otherwise, in TDD, the fixed-uplink and fixed-downlink designs perform almost as well as the individual design upper bound, and thus reduce complexity, overhead and configuration periods with almost no loss in spectral efficiency.}}
\end{itemize}
\subsubsection*{Organization} In Section \ref{sec 2}, the system model and performance metrics are provided. In Section \ref{sec 3}, the weighted-sum problem (WSP) of jointly maximizing the DL-WSR and UL-WSR is formulated. The WSP is optimized using the block-coordinate descent (BCD) algorithm developed in Section \ref{sec 4}.  \textcolor{black}{Numerical experiments that show the efficacy of the proposed joint design, fixed designs and slicing benchmarks are presented in Section \ref{sec 5}.} \textcolor{black}{ Section \ref{sec 5} also includes a complexity analysis and discusses the trade-off between the three designs.} Conclusions are presented in Section \ref{sec 6}. 
\begin{figure*}%
\centering
\begin{subfigure}{.8\columnwidth}
\captionsetup{justification=centering,margin=1cm}
 \vspace{0.6cm}
\centerline{ \includegraphics[width=\linewidth]{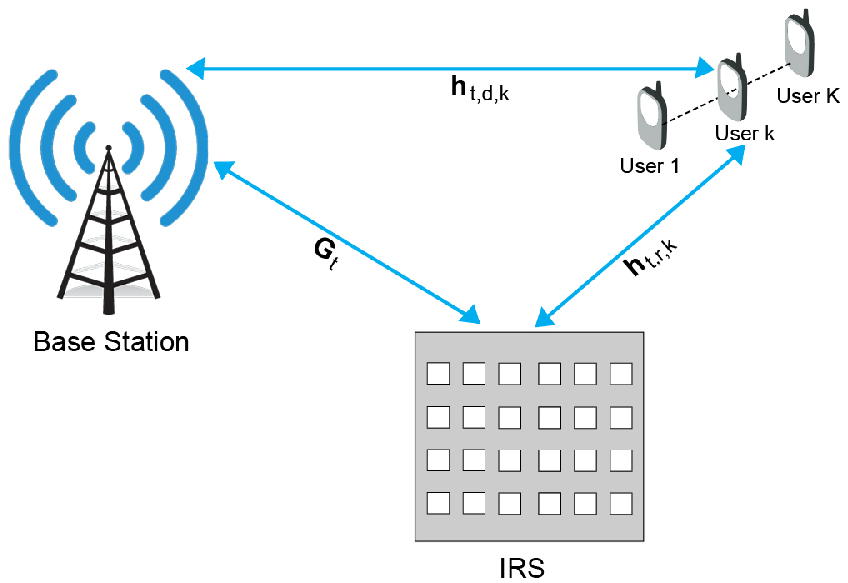}}
\vspace{0.5cm}
\label{fig:layoutt}
\subcaption{System diagram}
\end{subfigure} \hspace{2cm}
\begin{subfigure}{.8\columnwidth}
 \captionsetup{justification=centering,margin=1cm}
\centerline{ \includegraphics[width=\linewidth]{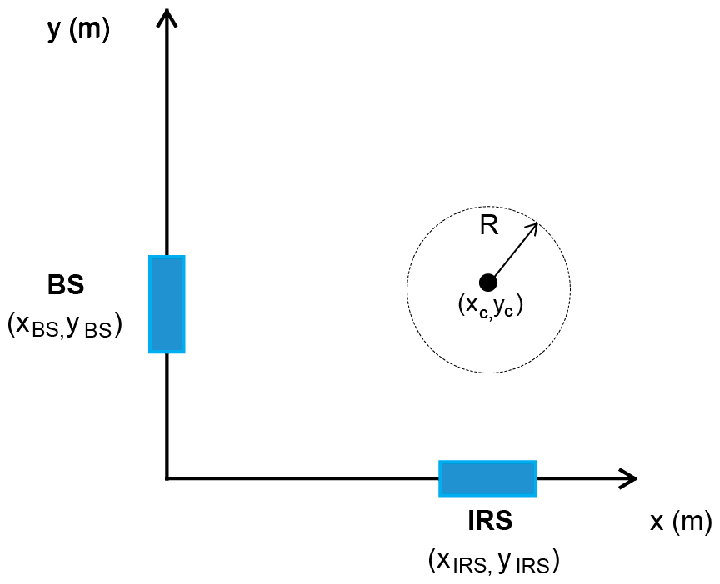}}
\subcaption{Top-view layout}
\end{subfigure}
 \caption{\textcolor{black}{(a) System diagram and (b) top-view layout for MU-MISO IRS assisted system.}}
 \label{fig:System layout}
\end{figure*}
\subsubsection*{Notation} The sets of real and complex numbers are denoted by $\mathbb{R}$ and $\mathbb{C}$. Moreover, $\mathcal{R}(.)$ and $\mathcal{I}(.)$ are used to represent the real and the imaginary parts of a complex variable, respectively. The transpose, conjugate and conjugate transpose operator are denoted using the superscripts $(.)^T$ , $(.)^*$ and $(.)^H$, respectively. Vectors are denoted using lower case bold symbols (e.g $\boldsymbol{v}$) and matrices are denoted using upper case bold symbols (e.g $\boldsymbol{M}$). For a vector $\boldsymbol{v} \in \mathbb{C}^N$,  $diag(\boldsymbol{v})$ is the $N \times N$ diagonal matrix whose diagonal entries are filled with the $N$ elements of $\boldsymbol{v}$.   The expectation of a random variable (RV) is denoted by $\mathbb{E} \left\{ . \right\}$ and $\mathcal{CN}(0,\sigma^2)$ is used to denote a circularly symmetric complex Gaussian RV with a mean of 0 and variance $\sigma^2$.
\section{System Model and Performance metrics}\label{sec 2}
\subsection{Model}
In this paper, a single-cell communications system is considered where an $M$-antenna BS is assisted by an $N$-element IRS in order to serve $K$ single-antenna users. The users are indexed by $\mathcal{K}=\{1,....,K\}$. The system diagram and layout are shown in Fig.~\ref{fig:System layout}. 
\par
In the DL, the precoded signal transmitted by the BS is expressed as $\boldsymbol{x^{DL}} = \sum_{k=1}^{K} {\boldsymbol{w_k} s^{DL}_k}$ where $\boldsymbol{w_k}\in\mathbb{C}^{M \times 1 }$ and $s^{DL}_k\in\mathbb{C}$ are the downlink transmit beamforming vector and  the information symbol in the downlink associated with user $k\in\mathcal{K}$, respectively. \textcolor{black}{The information symbols are zero-mean and satisfy  $\mathbb{E}\{s_k^{DL} \left(s_{i}^{DL}\right)^H\} = \delta_{i,k}$, i.e., are i.i.d. with unit power.} 
\par
In the UL, the signal transmitted by user $k$ can be expressed as $x^{UL} = \sqrt{p_k} s^{UL}_{k}$ where $p_k>0$ and $s^{UL}_k \in \mathbb{C}$ are the uplink power transmitted by user $k$ and the information symbol in the uplink associated with user $k$, respectively. The UL information symbols satisfy $\mathbb{E}\{s_k^{UL} \left(s_{i}^{UL}\right)^H\} = \delta_{i,k}$. Additionally, $\boldsymbol{v_k} \in \mathbb{C}^{M \times 1 }$ is the unit-norm receive combining vector at the BS for user $k$. 
\par 
The elements of the IRS are indexed by $\mathcal{N}=\{1,\ldots,N\}$ with the vector containing the IRS phase shifts expressed as   $\boldsymbol{\theta}= [\theta_1,.....,\theta_N]$. For each IRS element $n \in \mathcal{N}$, the reflection coefficient is represented as $\theta_n = e^{-j\varphi_n }$ where $\varphi_n \in [0,2\pi)$ is the phase shift associated with element $n$. The phase-shift matrix associated with the IRS can then be expressed as $\boldsymbol{\Theta}= diag(\boldsymbol{\theta})$.  
\par
All channels are quasi-static flat and it is assumed that channel state information (CSI) is perfectly known. Let $\boldsymbol{h}_{t,d,k}$ $\in$ $\mathbb{C}^{M \times 1 }$ denote the baseband equivalent direct channel between the BS and user $k$, $\boldsymbol{h}_{t,r,k}$ $\in$ $\mathbb{C}^{N \times 1 }$ denote the channel between the IRS and user $k$, and  $\boldsymbol{G}_{t}$ $\in$ $\mathbb{C}^{M \times N }$ denote the direct channel between the BS and IRS, where $t \in {\{UL,DL\}}$ indicates whether the uplink channel or the downlink channel is considered. In TDD systems, the downlink and uplink transmissions occur over the same  frequency bands but at different time slots. Moreover, channel reciprocity in TDD, as experimentally established for IRS-assisted systems, implies that the uplink and downlink channels are the same \cite{IRS-reciporcity-1,reciprocity-2}. In FDD, the downlink and uplink transmissions occur concurrently over different frequency bands, thus the channels are generally different for uplink and downlink transmissions.
\par 
The effective channels between the BS and user $k$ in DL and UL are expressed as  
\begin{subequations}
     \begin{align}
     \boldsymbol{h}_{DL,k}&= \boldsymbol{h}_{DL,d,k} + \boldsymbol{G}_{DL} \, \boldsymbol{\Theta}_{DL} \, \boldsymbol{h}_{DL,r,k} \label{1a}    \\
      \boldsymbol{h}_{UL,k}&= \boldsymbol{h}_{UL,d,k} + \boldsymbol{G}_{UL} \,  \boldsymbol{\Theta}_{UL} \,
      \boldsymbol{h}_{UL,r,k} \,. \label{1b} 
    \end{align}   
\end{subequations}
\par
\textcolor{black}{It is important to note that in general TDD systems, the IRS configuration $\boldsymbol{\Theta}$ can be different for UL and DL transmissions, {although at the cost of increased overhead, power consumption and {configuration periods}}, whereas for FDD the same IRS configuration is used to support simultaneous UL-DL transmissions.} \textcolor{black}{In this work, a  joint IRS configuration is used for UL and DL transmissions. However, the case where the IRS configurations can be different for UL and DL transmissions is also considered to evaluate the performance of the joint design.}  
\subsection{Linear beamforming and performance metrics}
\par 
The signal received by user $k$ in the DL is 
\begin{align}
        y_{k}^{DL} &= \underbrace{\vphantom{\left(\sum_{i=1, i \neq k}^{K} (\boldsymbol{h}_{DL,k})^{\textcolor{black}{T}} \boldsymbol{w}_i  s_i^{DL}\right)}
    (\boldsymbol{h}_{DL,k})^{\textcolor{black}{T}} \boldsymbol{w}_k s_k^{DL}}_{\text{Intended signal}} + \underbrace{\sum_{i=1, i \neq k}^{K} (\boldsymbol{h}_{DL,k})^{\textcolor{black}{T}} \boldsymbol{w}_i s_i^{DL}}_{\text{MU-interference}} \nonumber \\  &  + \underbrace{n_k^{DL},}_{\text{AWGN noise}} 
\end{align}
where $n_k^{DL} \sim \mathcal{CN} (0,\sigma_{DL}^2)$ is  circularly symmetric complex additive white Gaussian noise (AWGN) at user $k$.   Similarly, the signal received by the BS in UL is expressed as 
\begin{equation}
    \boldsymbol{y}^{UL} =\sum_{i=1}^{K} \sqrt{p_i}\, \boldsymbol{h}_{UL,i}\, s_i^{UL}.
\end{equation}
If $\boldsymbol{v}_k$ is the combining vector at the BS for user $k$, the combined signal is given by
\begin{multline}
\label{eqn:5}
   \boldsymbol{v_k}^H\,\boldsymbol{y}^{UL} =\underbrace{\sqrt{p_k} \,\boldsymbol{v_k}^H \,\boldsymbol{h}_{UL,k} \,s_k^{UL}}_{\text{Intended signal}} \\
   +   \underbrace{\boldsymbol{v_k}^H \left( \sum_{i=1, i \neq k}^{K} \sqrt{p_i}\, \boldsymbol{h}_{UL,i}\, s_i^{UL} \right)}_{\text{MU-interference}} + \underbrace{\vphantom{ \left(\sum_{i=1, i \neq k}^{K} (\boldsymbol{h}_{DL,k})^{H} \boldsymbol{w}_i \, s_i^{DL}\right) }\boldsymbol{v_k}^H \left( \boldsymbol{n}_k^{UL} \right)}_{\text{AWGN noise}},
\end{multline}
where $\boldsymbol{n}_k^{UL} \sim \mathcal{CN} (0,\sigma_{UL}^2 \mathbb{I}_M)$ is the AWGN at the BS.
\par
The SINR corresponding to user $k$ in DL and UL are given by
\begin{subequations} 
\begin{align}
\label{eqn:6}
\gamma_{k}^{DL} &= \frac{ |\left(\boldsymbol{h}_{DL,k}\right)^{\textcolor{black}{T}} \boldsymbol{w}_{k}|^2}{ \sigma_{DL}^2  +\sum_{i=1, i\neq k}^K |\left(\boldsymbol{h}_{DL,k}\right)^{\textcolor{black}{T}} \boldsymbol{w}_{i}|^2} \\
\label{eqn:7}
 \gamma_{k}^{UL} &= \frac{p_{k} | \left(\boldsymbol{h}_{UL,k}\right)^H \boldsymbol{v}_{k}|^2}{\boldsymbol{v}_{k}^{H} \left(\sigma_{UL}^2 \; \boldsymbol{I}_M +\sum_{i=1, i\neq k}^K p_{i}\, \boldsymbol{h}_{UL,i} \left(\boldsymbol{h}_{UL,i}\right)^H  \right)\boldsymbol{v}_{k} }. 
\end{align}
\end{subequations}
\section{{Problem Formulation}} \label{sec 3}
\textcolor{black}{In this section, we formulate the IRS-assisted UL and DL weighted sum-rate problems for FDD and TDD systems. Then, in order to obtain {a trade-off region between uplink and downlink rates achieved by a joint IRS design, a weighted sum problem is formulated.}}
\subsection{Problem Formulation}
In DL and UL, the achievable data rates in bit/sec/Hz associated with user $k$ are given by 
\begin{subequations}
    \begin{align}
    \Tilde{R}_k^{DL} &= \log_2(1+ \gamma_k^{DL}) \qquad \text{[bit/sec/Hz]} \\
    \Tilde{R}_k^{UL} &= \log_2(1+ \gamma_k^{UL}) \qquad \text{[bit/sec/Hz]}.
    \end{align}
\end{subequations} 
Asymmetrical demands for UL vs DL transmission arise in practical  communications systems. In TDD systems, flexible resource allocation becomes feasible by adjusting the proportion of time resources dedicated for UL vs DL.
Rather than assigning equal resources to UL and DL, a normalized weight, $\alpha \in [0,1]$, is introduced in order to capture the relative proportion of time resources dedicated to DL vs UL transmissions in TDD systems. 
\par
In a TDD system with \textcolor{black}{bandwidth $B$ Hz}, the DL/UL rate in bit/sec is then given by 
\begin{subequations}
    \begin{align}
    \label{10a}
    \bar{R}_k^{DL} &=  \alpha \; B \log_2(1+ \gamma_k^{DL}) \quad \; \; \;  \; \; \qquad \text{[bit/sec]} \\
    \label{10b}
    \bar{R}_k^{UL} &= (1-\alpha) \; B \log_2(1+ \gamma_k^{UL}) \qquad \text{[bit/sec]}
    \end{align}
\end{subequations}
where $\alpha$ is the fraction of time used for DL. 
Normalizing \eqref{10a} and \eqref{10b} by $B$, we have
\begin{subequations}
    \begin{align}
    \label{*}
    {R}_k^{DL} &=  \alpha \; \log_2(1+ \gamma_k^{DL}) \quad  \quad \quad   \; \; \; \; \;  \text{[bit/sec/Hz]} \\
    \label{**}
    {R}_k^{UL} &= (1-\alpha)  \log_2(1+ \gamma_k^{UL}) \; \quad  \quad  \text{[bit/sec/Hz]}.    \end{align}
\end{subequations}
Likewise, in an FDD system with a bandwidth of \textcolor{black}{$\alpha B$ dedicated to downlink and $(1-\alpha) B$}
dedicated to the uplink, the spectral efficiency is again given by \eqref{*} and \eqref{**} respectively. 
\par
  The first objective is to maximize the weighted sum-rate in DL by jointly optimizing the phase-shifts at the IRS and the transmit beamforming vectors, $\boldsymbol{W}:= \{\boldsymbol{w_1},\ldots,\boldsymbol{w_k}\}$, at the BS which are subject to a sum-power constraint. The aforementioned problem is given by
  \begin{subequations}
 \label{eq:DL_WSR}
\begin{align}
& \underset{\boldsymbol{\Theta}_{DL}, \boldsymbol{W}}{\text{max}}
& & J_{DL}= \sum_{k=1}^{K} \varepsilon_k^{DL}
R_k^{DL}  \\
\label{DL_const}
& \text{subject to}
& & \sum_{k=1}^{K} \| \boldsymbol{w_k}\|^2 \leq P_{max}^{DL}, \\
&&& |[\boldsymbol{\Theta}_{DL}]_{n,n}| = 1, \qquad \qquad n = 1, \ldots, N. 
\end{align}
\end{subequations}
where $\varepsilon_k^{DL}>0$ is the weight associated with user $k$ in DL, and captures the relative priority of user $k$ in the DL. Moreover, $P_{max}^{DL}$ denotes the maximum sum-power at the BS.
\par
In TDD, $P_{max}^{DL} = B S_{max}^{DL}$ where $S_{max}^{DL}$ is the maximum downlink power spectral density in Watts/Hz. \textcolor{black}{Similarly, the noise power in downlink is expressed as $\sigma_{DL}^2 = B N_0$ where $N_0$ is the noise power spectral density. In FDD, \textcolor{black}{$P_{max}^{DL} = \alpha B S_{max}^{DL}$ and  $\sigma_{DL}^2 = \alpha B N_0$.}}
\par
\textcolor{black}{The second objective is to maximize the weighted sum-rate in UL by jointly optimizing the phase-shifts at the IRS, the unit-norm receive beamforming vectors at the BS defined by $\boldsymbol{V}:=\{\boldsymbol{v_1},\ldots,\boldsymbol{v_k}\}$ and the uplink power control $\boldsymbol{P}:=[p_1,\ldots,p_K]$ where each $p_k$ is subject to a max-power constraint $P_{max}^{UL}$. Maximizing the WSR in UL is expressed as}
\begin{subequations}
\label{eq:UL_WSR}
\begin{align}
& \underset{\boldsymbol{\Theta}_{UL},\boldsymbol{V},\boldsymbol{P}}{\text{max}}
& & J_{UL}= \sum_{k=1}^{K} \varepsilon_k^{UL}
R_k^{UL} \\
\label{UL_const}
& \text{subject to}
& &  0 \leq p_k \leq P_{max}^{UL}, \qquad  k = 1, \ldots, K. \\
&&& |[\boldsymbol{\Theta}_{UL}]_{n,n}| = 1, \qquad  n = 1, \ldots, N. 
\end{align}
\end{subequations}
where $\varepsilon_k^{UL} > 0$ is the weight associated with user $k$ in UL that captures the relative priority of user $k$ in UL. Moreover, $P_{max}^{UL}$ denotes the maximum power that can be transmitted by a user. In TDD, $P_{max}^{UL} = B S_{max}^{UL}$ where $S_{max}^{UL}$ is the maximum uplink power spectral density in Watts/Hz. \textcolor{black}{Additionally, the noise power in uplink is $\sigma_{UL}^2 = B N_0$. In FDD, \textcolor{black}{$P_{max}^{UL} = (1-\alpha) B S_{max}^{UL}$  and  $\sigma_{UL}^2 = (1-\alpha) B N_0$.}}
\par 
For a joint IRS design, problems \eqref{eq:DL_WSR} and \eqref{eq:UL_WSR} are coupled through the IRS phase shifts $\boldsymbol{\Theta}_{DL}=\boldsymbol{\Theta}_{UL} :=\boldsymbol{\Theta}$. The multi-objective optimization problem is scalarized by formulating a weighted sum problem (WSP) by introducing a weight $\beta \in [0,1]$ that captures the relative priority given to DL vs UL when optimizing the IRS.
The WSP is then formulated as
\begin{subequations}
\label{eq:UL_DL_WSR}
\begin{align}
 \underset{\boldsymbol{\Theta},\boldsymbol{V},\boldsymbol{W},\boldsymbol{p}}{\text{max}} 
 J_{WSP}=& \alpha \beta  \sum_{k=1}^{K} \varepsilon_k^{DL} \log_2( 1+ \gamma_k^{DL}) \\ \nonumber &+  (1-\alpha)(1-\beta)  \sum_{k=1}^{K} \varepsilon_k^{UL}
\log_2(1+ \gamma_k^{UL})\\
 \text{subject to} \; \; 
\eqref{DL_const}&,\eqref{UL_const}
 \\
 \quad \quad  |[\boldsymbol{\Theta}&]_{n,n}|   = 1,  \quad  n = 1, \ldots, N.
\end{align}
\end{subequations}
\subsection{UL/DL user weighting strategies}
\label{sec: user weighting}
\par
The weighting strategy of the UL/DL user is an important design parameter that can be chosen to prioritize users. Three weighting strategies are considered: \emph{equal}, \emph{proportional-fair} and \emph{independent}, and the weights are always normalized such that $\sum_{k=1}^{K} \varepsilon_k^{UL} = 1$ and $\sum_{k=1}^{K} \varepsilon_k^{DL} = 1$. 
\par
Under the {equal} weight strategy, $\varepsilon_k^{UL} = \varepsilon_k^{DL} = \frac{1}{|\mathcal{K}|}$ which corresponds to providing uniform priority to the users. In order to provide further flexibility in assigning resources while maintaining fairness, the weights in \eqref{eq:UL_DL_WSR} can also be chosen based on a {proportional-fair} (PF) strategy. In particular, memory is incorporated into the system in order to capture the history of the data transmissions provided for each user in the past. Let $r_{k,t}^{s}$ be the data rate associated with user $k$ at time slot $s$ in the DL or the UL indicated by $t \in \{UL,DL\}$. The objective associated with PF \textcolor{black}{at time-slot $s$ is}  \cite{interference_book,memory_PF}
\begin{multline*}
     \max_{\textcolor{black}{\{r_{k,t}^{s}\}_{k=1}^K}}  \prod_{k=1}^K \left(r_{k,t}^{s} + \sum_{l < s} r_{k,t}^{l}\right)  \iff \\  \quad \quad \max_{\textcolor{black}{\{r_{k,t}^{s}\}_{k=1}^K}} \sum_{k=1}^K \log \left(1+\frac{r_{k,t}^{s}}{\sum_{l < s} r_{k,t}^{l}}\right) +\sum_{k=1}^K \log \left( \sum_{l < s} r_{k,t}^{l} \right).
\end{multline*}
\textcolor{black}{Then for $\sum_{l < s} r_{k,t}^{l} \gg  r_{k,t}^{s} $, approximating $\log (1+x) \approx x$ and by ignoring $\sum_{k=1}^K \log \left( \sum_{l < s} r_{k,t}^{l} \right)$ (which is a constant for each slot), the weights in \eqref{eq:UL_DL_WSR} are given \textcolor{black}{by $\varepsilon_k^{t}= \frac{\Gamma_t}{\sum_{l < s} r_{k,t}^{l}}$ where $\Gamma_t$ is a normalization constant chosen such that $\sum_{k=1}^{K} \varepsilon_k^{t} = 1$ .}}
\textcolor{black}{Finally, we also consider a scenario where the weights are uncoupled from user channel quality and position. Specifically, we consider {independent} user-weights where the weights are randomly chosen independently of all user/network parameters and then normalized.} This can correspond to a scenario where the demand for downlink and uplink transmissions vary substantially across the users and the variance of the weights reflect the flexibility needed to accommodate the requests for transmissions.
\section{\textcolor{black}{Optimizing The Joint Design, and Alternate Schemes}} 
\label{sec 4}
\par 
\textcolor{black}{To tackle the non-convex \textcolor{black}{weighted sum} optimization problem in \eqref{eq:UL_DL_WSR}, a block-coordinate descent (BCD) algorithm is adopted. We iteratively leverage RCG, WMMSE and fractional programming (FP) in a BCD algorithm in order to efficiently {optimize} the WSP formulated in \eqref{eq:UL_DL_WSR}. {The coupled optimization variables of the WSP in \eqref{eq:UL_DL_WSR} are decomposed  into blocks of decoupled sub-problems that are iteratively optimized.}}
\par
 \textcolor{black}{ Specifically,  for a fixed IRS configuration $\boldsymbol{\Theta}$, the WSP in \eqref{eq:UL_DL_WSR} decouples into separate downlink  and uplink sub-problems. The WMMSE algorithm is used to optimize the transmit beamforming vectors $\{\boldsymbol{w}_k\}_{k=1}^K$ of the downlink sub-problem. For fixed $\boldsymbol{\Theta}$ and uplink power $\{p_k\}_{k=1}^K$, an MMSE filter is used to optimize the receive beamforming vectors $\{\boldsymbol{v}_k\}_{k=1}^K$ and fractional programming (FP) is used to optimize  the uplink power $\{p_k\}_{k=1}^K$ for fixed $\boldsymbol{\Theta}$ and $\{\boldsymbol{v}_k\}_{k=1}^K$. Moreover, for fixed $\{\boldsymbol{w}_k\}_{k=1}^K$, $\{\boldsymbol{v}_k\}_{k=1}^K$ and $\{p_k\}_{k=1}^K$, RCG is used to update $\boldsymbol{\Theta}$. The algorithm  proceeds by iterating over the decoupled blocks and updating the decision variables of a particular block while fixing the variables associated with the other blocks.}
\subsection{Updating the beamforming vectors and power control for fixed IRS configuration}
\par 
\textcolor{black}{For fixed $\boldsymbol{\Theta}$, the WSP in \eqref{eq:UL_DL_WSR} decouples into a separate downlink sub-problem in \eqref{eq:DL_WSR} and uplink sub-problem in \eqref{eq:UL_WSR}.}
\subsubsection{Updating transmit beamforming vectors $\{\boldsymbol{w}_k\}_{k=1}^K$} 
\textcolor{black}{when $\boldsymbol{\Theta}$ is fixed, optimizing the transmit beamforming vectors in \eqref{eq:DL_WSR} corresponds to a DL-WSR maximization problem. {{As in \cite{WMMSE},}} by introducing the auxiliary variables $\{u_k\}_{k=1}^K$ and  $\{t_k\}_{k=1}^K$ and  with $\boldsymbol{u}= [u_1,\ldots, u_K]$, $\boldsymbol{t}= [t_1,\ldots, t_K]$, the DL-WSR problem in \eqref{eq:DL_WSR} is equivalent to the WMMSE problem written as 
\begin{equation}
\begin{aligned}
\label{eq:DL_WMMSE}
& \underset{\boldsymbol{u},\boldsymbol{t},\boldsymbol{W}}{\text{min}}
& & \sum_{k=1}^{K} \varepsilon_k^{DL} \left( u_k e_k - \log_2 u_k \right)  \\
& \text{subject to}
& & \sum_{k=1}^{K} \| \boldsymbol{w_k}\|^2 \leq P_{max}^{DL}, \\
\end{aligned}
\end{equation}
where the MSE associated with user $k$ is $e_k = \mathbb{E}_{\boldsymbol{s},n_k^{DL}} \left\{(s^{DL}_k - \hat{s}^{DL}_{k}) (s^{DL}_k - \hat{s}^{DL}_k)^H \right\}$
where $\boldsymbol{s}= [s_{1}^{DL},\ldots,s_{K}^{DL}]$ and the estimated DL symbol at user $k$ is given by $\hat{s}^{DL}_k= t_k \; y_k^{DL}$.} 
\par 
\textcolor{black}{The equivalent WMMSE problem lends itself to a tractable iterative algorithm as the problem in \eqref{eq:DL_WMMSE} is convex in the optimization variables when the other variables are fixed to the values of their previous iteration.
\noindent Proceeding similar to \cite{WMMSE}, the problem in \eqref{eq:DL_WMMSE} is optimized by iterating over the following update rules for each user $k \in \mathcal{K}$:
\begin{subequations}
\label{WMMSE_updates}
\begin{align}
\label{a}
u_k &= \frac{\sum_{i=1}^{K} \left| \boldsymbol{h}_{DL,k}^{{T}} \; \boldsymbol{w}_i  \right|^2 + \sigma_{DL}^2}{\sum_{i=1, i \neq k }^{K} \left| \boldsymbol{h}_{DL,k}^{{T}} \; \boldsymbol{w}_i  \right|^2 + \sigma_{DL}^2}\\
\label{b}
t_k &= \frac{\boldsymbol{h}_{DL,k}^{{T}} \; \boldsymbol{w}_k}{\sum_{i=1}^{K} \left| \boldsymbol{h}_{DL,k}^{{T}} \;  \boldsymbol{w}_i  \right|^2 + \sigma_{DL}^2}\\
\label{c}
\boldsymbol{w}_k &= \varepsilon_k^{DL} u_k t_k {\Big (}\nu \; \boldsymbol{I}_M \\ & \quad +  \sum_{i=1}^{K}  \varepsilon_i^{DL} u_i |t_i|^2 \boldsymbol{h}_{DL,i}^{{*}} \;  \boldsymbol{h}_{{DL,i}}^{{T}}  {\Big )}^{-1} \boldsymbol{h}_{DL,k}^{{*}} \nonumber,
\end{align}
\end{subequations}
and $\nu \geq 0$ is the optimal Lagrange multiplier chosen such that complementary slackness associated with the sum-power constraint is achieved. By iterating over  \eqref{a}, \eqref{b} and \eqref{c}, {Theorem 3} of \cite{WMMSE} guarantees convergence to a stationary point of the  \eqref{eq:DL_WMMSE}.}
\subsubsection{Updating receive beamforming vectors $\{\boldsymbol{v}_k\}_{k=1}^K$} \textcolor{black}{when $\boldsymbol{\Theta}$ and $\{p_k\}$ are fixed, the objective in the uplink sub-problem of \eqref{eq:UL_WSR}, is maximized by the MMSE receiver. For each user $k \in \mathcal{K}$, the receive beamforming vector is given as
\begin{equation}
\label{eq:UL_MMSE}
    \begin{aligned}
     \boldsymbol{v}_k = \frac{\left(\sigma_{UL}^2   \boldsymbol{I}_M + \sum_{i=1}^{K} p_i   \boldsymbol{h}_{UL,i}  \boldsymbol{h}_{UL,i}^H  \right)^{-1} \boldsymbol{h}_{UL,k}} {\left \lVert \left(\sigma_{UL}^2   \boldsymbol{I}_M + \sum_{i=1}^{K} p_i   \boldsymbol{h}_{UL,i}  \boldsymbol{h}_{UL,i}^H  \right)^{-1} \boldsymbol{h}_{UL,k} \right \rVert }.
    \end{aligned}
\end{equation}}
\subsubsection{Updating uplink power control $\{p_k\}_{k=1}^K$} \textcolor{black}{when $\boldsymbol{\Theta}$ and $\{\boldsymbol{v}_k\}_{k=1}^K$ are fixed, fractional programming (see \cite{FP}, \cite{FP_2}) is used to optimize $\{p_k\}_{k=1}^K$ in \eqref{eq:UL_WSR}. The Lagrangian dual reformulation is applied by introducing the auxiliary variable $\boldsymbol{\gamma} = [\gamma_1,\ldots,\gamma_K]$. The power control problem of  \eqref{eq:UL_WSR} is equivalent to:
\begin{align}
 \label{eq:PC_LDT}
   \underset{\boldsymbol{\gamma},\boldsymbol{P}}{\text{max}} \quad 
  \sum_{k=1}^{K} \varepsilon_k^{UL} &
 \ln(1+\gamma_k)  -\sum_{k=1}^{K} \varepsilon_k^{UL}
 \gamma_k \nonumber \\  \quad + & \sum_{k=1}^{K} \frac{\varepsilon_k^{UL}
 (1+\gamma_k) \left|\boldsymbol{v}_k^H\boldsymbol{h}_{UL,k}  \right|^2 p_k}{\sum_{i=1}^{K} \left|\boldsymbol{v}_k^H\boldsymbol{h}_{UL,i} \right|^2 p_i + \sigma_{UL}^2 },   \\ \nonumber
    \text{subject to} \quad \quad  
    &0  \leq  p_k \leq P_{max}^{UL}, \quad   k = 1, \ldots, K.
\end{align}  
For fixed $\{p_k\}_{k=1}^K$, setting the derivative of the objective in \eqref{eq:PC_LDT} with respect to $\gamma_k$ to zero and solving for $\gamma_k$, yields $\gamma^{\star}_k$ for each $k \in \mathcal{K}$ which is given by
 \begin{equation}
     \begin{aligned}
     \label{gamma update}
     \gamma^{\star}_k = \frac{ \left|\boldsymbol{v}_k^H\boldsymbol{h}_{UL,k}  \right|^2 p_k}{\sum_{i=1,i\neq k}^{K} \left|\boldsymbol{v}_k^H\boldsymbol{h}_{UL,i} \right|^2 p_i + \sigma_{UL}^2 }.
     \end{aligned}
 \end{equation} 
By applying the quadratic transform  and proceeding similar to \cite{FP}, the power control update for each user $k$ is obtained as
\begin{equation}
    \begin{aligned}
    \label{power update}
    p_k^{\star} = \min \left\{P_{max}^{UL} \; , \; \frac{\chi_k^2 \; \varepsilon_k^{UL}
 (1+\gamma_k) \left|\boldsymbol{v}_k^H\boldsymbol{h}_{UL,k}  \right|^2}{\left(\sum_{i=1}^{K} \chi_i^2 \left|\boldsymbol{v}_i^H\boldsymbol{h}_{UL,k} \right|^2 \right)^2}\right\},
    \end{aligned}
\end{equation}
where $\gamma_k$ is obtained using  \eqref{gamma update} and the auxiliary variables $\{\chi_k\}$ are updated as:
\begin{equation}
    \begin{aligned}
    \label{chi update}
    \chi_k^{\star} =  \frac{ \sqrt{\; \varepsilon_k^{UL}
 (1+\gamma_k) \left|\boldsymbol{v}_k^H\boldsymbol{h}_{UL,k}  \right|^2 p_k} }{\sum_{i=1}^{K}  \left|\boldsymbol{v}_k^H\boldsymbol{h}_{UL,i} \right|^2 p_i + \sigma_{UL}^2}.
    \end{aligned}
\end{equation}
Iterating over the updates in \eqref{chi update}, \eqref{gamma update} and \eqref{power update} yields a fixed point of the power control problem in \eqref{eq:PC_LDT} as established in \cite{FP_2}.}
\subsection{Updating the IRS configuration for fixed beamforming vectors and uplink power control}
\textcolor{black}{For fixed $\{\boldsymbol{w}_k\}_{k=1}^K$, $\{\boldsymbol{v}_k\}_{k=1}^K$ and $\{p_k\}_{k=1}^K$, ignoring terms that don't depend on $\boldsymbol{\theta}$, the WSP in \eqref{eq:UL_DL_WSR} \textcolor{black}{is equivalent} to:
\begin{align}
    \label{eq:UL_DL_MO}
  \underset{\boldsymbol{\theta}}{\text{max}} \quad J_{MO}(\boldsymbol{\theta}) = \alpha  &\beta  \sum_{k=1}^{K} \varepsilon_k^{DL} R_{MO,k}^{DL}(\boldsymbol{\theta})
\\ \nonumber +& (1-\alpha)   (1-\beta)  \sum_{k=1}^{K} \varepsilon_k^{UL}
R_{MO,k}^{UL}(\boldsymbol{\theta})\\ \nonumber 
\text{subject to} 
\;  |\theta_n| = 1&,    \; n = 1, \ldots, N
\end{align}
Here for \textsf{$t\in \{DL,UL\}$}, $R_{MO,k}^{t}{(\boldsymbol{\theta})}$ is given as 
\begin{equation}
    \begin{aligned}
    R_{MO,k}^{\;t}(\boldsymbol{\theta})&= \log \left( 1+ \frac{\left|\boldsymbol{\theta}^H \boldsymbol{\lambda}_{t,k,k} + \mu_{t,k,k}  \right|^2}{\sum_{i=1,i\neq k}^{K} \left|\boldsymbol{\theta}^H \boldsymbol{\lambda}_{t,i,k} +  \mu_{t,i,k}\right|^2 + \sigma_{k,t}^2} \right),
    \end{aligned}
\end{equation}
where
\begin{subequations}
    \begin{align}
    \boldsymbol{\lambda}_{DL,i,k} &=  \text{diag}\left(\boldsymbol{h}_{DL,r,k}^{{H}}\right) \boldsymbol{G}^{{H}}_{DL} \boldsymbol{w}_i^*, 
    \label{21a} \\ 
    \boldsymbol{\lambda}_{UL,i,k} &= \textcolor{black}{\sqrt{p_i}} \;  \text{diag}\left(\boldsymbol{h}_{UL,r,i}^{H}\right) \boldsymbol{G}^H_{UL} \boldsymbol{v}_k,  \nonumber \\
    \mu_{DL,i,k} &= \boldsymbol{h}_{DL,d,k}^{{H}} \boldsymbol{w}_i^*,
     \label{21b} \\
    \mu_{UL,i,k} &= \textcolor{black}{\sqrt{p_i}} \; \boldsymbol{h}_{UL,d,i}^H \boldsymbol{v}_k. \nonumber
    \end{align}
\end{subequations}
The $|\theta_n|=1$ constraint of the optimization problem in \eqref{eq:UL_DL_MO} defines a search space characterized by a  product of $N$ complex circles, thus representing a Riemannian submanifold of $\mathbb{C}^N$. Moreover, the function $J_{MO}(\boldsymbol{\theta})$ is differentiable. Consequently, a stationary point of the problem in \eqref{eq:UL_DL_MO} can be obtained using the Riemannian conjugate gradient (RCG) algorithm \cite{IRS-MO,early_IRS_MO}.}
\par 
\textcolor{black}{For any point $\boldsymbol{\theta}_{j}$ on a manifold $\mathcal{M}$, the tangent space denoted as $\mathcal{T}_{\boldsymbol{\theta}_j}\mathcal{M}$ comprises all the tangent vectors, each of which defines a search direction that can be used to optimize an objective function. The Riemannian gradient, denoted as $\text{grad}_{\boldsymbol{\theta}_j} J_{MO}$, represents the direction along which the objective function experiences the steepest increase and in the case of the complex circle manifold (CCM) of \eqref{eq:UL_DL_MO}, is given by
\begin{equation}
    \begin{aligned}
    \label{eq:Rgrad}
    \text{grad}_{\boldsymbol{\theta}_j} J_{MO} &= \nabla J_{MO} - \mathcal{R}\{\nabla J_{MO} \circ    {\boldsymbol{\theta}_j^{*}}\} \circ {\boldsymbol{\theta}_j},   
    \end{aligned}
\end{equation}
with the Euclidean gradient $\nabla J_{MO}$ given as 
    \begin{align}
      \nabla J_{MO} &= \alpha  \beta  \sum_{k=1}^K 2 \varepsilon_k^{DL} \boldsymbol{\Lambda}_{DL,k}  \nonumber \\ &+  (1-\alpha) (1-\beta)  \sum_{k=1}^K 2 \varepsilon_k^{UL} \; \boldsymbol{\Lambda}_{UL,k} , 
    \end{align}
where for $t \in \{DL,UL\}$, $\boldsymbol{\Lambda}_{t,k}$  is given by
\begin{align}
  & \boldsymbol{\Lambda}_{t,k} = \frac{\sum_{i=1}^K \boldsymbol{\lambda}_{t,i,k} \; \boldsymbol{\lambda}_{t,i,k}^H \; \boldsymbol{\theta}+ \sum_{i=1}^K \boldsymbol{\lambda}_{t,i,k}
     \; \mu_{t,i,k}^{*}}{\sum_{i=1}^K \left| \boldsymbol{\theta}^H \boldsymbol{\lambda}_{t,i,k} + \mu_{t,i,k}\right|^2 + \sigma_{t,k}^2 }  \label{eq: Gradients} \\ \nonumber
    & \qquad  -\frac{\sum_{i=1,i\neq k}^K \boldsymbol{\lambda}_{t,i,k} \; \boldsymbol{\lambda}_{t,i,k}^H \; \boldsymbol{\theta}+ \sum_{i=1,i \neq k}^K \boldsymbol{\lambda}_{t,i,k}
     \; \mu_{t,i,k}^{*}}{\sum_{i=1,i \neq k }^K \left| \boldsymbol{\theta}^H \boldsymbol{\lambda}_{t,i,k} + \mu_{t,i,k}\right|^2 + \sigma_{t,k}^2 }. 
\end{align}}
\par 
\textcolor{black}{For each iteration $j$, the RCG algorithm involves iterating over three updates:
\begin{itemize}
    \item[1-] Performing retraction by projecting the search direction $\boldsymbol{\eta}_{j}$  to the CCM using 
    \begin{equation}
    \label{MO-3}
    \boldsymbol{\theta}_{j+1} = \text{ret}_{CM}\left(\boldsymbol{\theta}_j + \alpha_{j}  \boldsymbol{\eta}_{j} \right),
    \end{equation}
    where $\alpha_{j}$ is an Armijo backtracking line search step size and  for a vector $\boldsymbol{x} \in \mathbb{C}^N$ $\text{ret}_{CM}\left(\boldsymbol{x}\right)$ is defined as
\begin{equation*}
    \begin{aligned}
    \text{ret}_{CM}\left(\boldsymbol{x}\right) = \left[\frac{x_1}{|x_1|},\ldots, \frac{x_N}{|x_N|}\right].
    \end{aligned}
\end{equation*}
    \item[2-] Finding the Riemannian Gradient $\text{grad}_{\boldsymbol{\theta}_{j+1}} J_{MO}$ according to \eqref{eq:Rgrad}   
    \item[3-] Finding the search direction  $\boldsymbol{\eta}_{j+1}$  by finding a tangent vector that is conjugate to $\text{grad}_{\boldsymbol{\theta}_{j+1}} J_{MO}$, given as 
    \begin{equation}
       \label{MO-1} \begin{aligned}
       \boldsymbol{\eta}_{j+1} = - \text{grad}_{\boldsymbol{\theta}_{j+1}} J_{MO} + \tau_j \mathcal{T}_{j,j+1} \left(  \boldsymbol{\eta}_{j} \right), \end{aligned}
    \end{equation}
    where $\tau_j$ is the Polak-Ribiere parameter obtained as in \cite{early_IRS_MO,MO-book} and the vector transport function $\mathcal{T}_{j,j+1}$ maps the tangent vector space at  $\boldsymbol{\theta}_{j}$  to the tangent vector space at $\boldsymbol{\theta}_{j+1}$. \\
    For the CCM of \eqref{eq:UL_DL_MO}, $\mathcal{T}_{j,j+1}$ is given by
    \begin{equation}
       \label{MO-2} \begin{aligned}
       \mathcal{T}_{j,j+1} \left(  \boldsymbol{\eta}_{j} \right) &= \boldsymbol{\eta}_{j} - \mathcal{R}\{\boldsymbol{\eta}_{j} \circ \boldsymbol{\theta}_{j+1}^{*}\} \circ \boldsymbol{\theta}_{j+1}  
        \end{aligned}
    \end{equation}
\end{itemize}
}

\textcolor{black}{The convergence of the RCG algorithm described  above to a critical point of the optimization problem in \eqref{eq:UL_DL_MO} is established in  \cite{MO-book}.}  \textcolor{black}{ Algorithm 1 summarizes the BCD algorithm used to obtain a fixed point of \eqref{eq:UL_DL_WSR}.}
\begin{algorithm}[t]
\caption{\textcolor{black}{BCD algorithm}}
\label{BCD algorithm}
\begin{algorithmic}[1]
\State \textcolor{black}{Initialize $\boldsymbol{\theta} = \boldsymbol{\theta}_1$, $\{\boldsymbol{v}_k\}_{k=1}^K$, $\{p_k\}_{k=1}^K$,  $\{\boldsymbol{w}_k\}_{k=1}^K$ and iteration $j=1$  
    \Repeat
  \State Update power $\{p_k\}_{k=1}^K$ using iterative algorithm \eqref{gamma update}, \eqref{power update} and \eqref{chi update}.
\State Update $\{\boldsymbol{v}_k\}_{k=1}^K$ using \eqref{eq:UL_MMSE}.
  \State Update $\{\boldsymbol{w}_k\}_{k=1}^K$ using iterative algorithm in \eqref{a}, \eqref{b} and \eqref{c}.
  \State Update $\boldsymbol{\theta}_j$ using RCG algorithm in \eqref{MO-3}, \eqref{MO-1} and \eqref{MO-2}.
\Until{$\lVert J_{WSP,j} - J_{WSP,j-1} \rVert < \epsilon$, $\epsilon>0$ is a threshold}}
\end{algorithmic}
\end{algorithm}
\subsection{Structure of beamforming vectors and channel reciprocity in IRS systems}
\label{3C}
 The joint IRS problem, {optimized using the developed BCD algorithm,} involves configuring the IRS phase shifts based on the downlink transmit beamforming and the uplink receive beamforming. \textcolor{black}{Consequently, a joint IRS design would benefit from any similarity in the structure of the uplink and downlink beamforming vectors.} Indeed, the similarity of the structure can be seen from the uplink beamforming vectors in \eqref{eq:UL_MMSE} and the downlink beamforming vectors in \eqref{WMMSE_updates}, as both beamforming vectors arise due to an MMSE filter with different weights and powers \textcolor{black}{(see \cite{UL_DL_duality}}). 
 \par
In TDD, channel reciprocity holds for the direct link {between the BS and the users as these links are} independent of the IRS and corresponds to a classical (non-IRS) system. Recently, experimental results in \cite{IRS-reciporcity-1} and \cite{reciprocity-2} have confirmed channel reciprocity for the cascaded link through the IRS. \textcolor{black}{Hence, for TDD operation, the joint IRS design would also benefit from  channel reciprocity and thus any similarity in the uplink and downlink beamforming vectors.} \textcolor{black}{Subsequently, the performance gap due to a joint design compared to individual design may be narrower for TDD compared to FDD. \textcolor{black}{Indeed, as will be shown in Section~\ref{sec 5}-B, for equal and proportional-fair weights, the joint design almost achieves the individual design bound in TDD, but not in FDD.} }
\subsection{Alternate Schemes}
\par
\textcolor{black}{For an IRS with $N$ elements, the {spectral efficiency} performance of the joint design in TDD/FDD systems is compared with optimizing the IRS for UL and DL transmissions individually and then the two IRS configurations are time-shared, which we refer to as \emph{Individual design}. Such a scheme is achievable for TDD but at the cost of increased overhead and {configuration periods} as explained before. However, this scheme is  unrealistic for FDD communications as both transmissions occur simultaneously.}  
\par 
\textcolor{black}{Moreover, the performance of the joint design is compared to two other designs called \emph{fixed-downlink} and \emph{fixed-uplink} design.} \textcolor{black}{In fixed-downlink design, the weighted sum rate (WSR) in \eqref{eq:DL_WSR} is first maximized by {jointly optimizing the IRS configuration $\boldsymbol{\Theta}$ and the transmit beamforming vectors} $\{\boldsymbol{w_k}\}_{k=1}^K$. Then, the receive combining vectors $\{\boldsymbol{v_k}\}_{k=1}^K$ are found using the effective channel {configured by the IRS phase shifts optimized using \eqref{eq:DL_WSR}}. In fixed-uplink design, the IRS configuration $\boldsymbol{\Theta}$  and the receive combining vectors  $\{\boldsymbol{v_k}\}_{k=1}^K$ {are first jointly optimized} to maximize the WSR in \eqref{eq:UL_WSR}. With this IRS configuration optimized for uplink transmission, the transmit beamforming vectors $\{\boldsymbol{w_k}\}_{k=1}^K$ are then found by applying a downlink beamforming strategy for the effective channel.}
\par 
\textcolor{black}{In addition, the performance of the joint design is also compared to two slicing benchmarks, namely: \emph{slicing-with-interference} and \emph{slicing-without-interference}. An $N$-element IRS is sliced by optimizing  $\frac{N}{2}$ IRS elements to exclusively assist each of the UL and DL transmissions. The slicing-without-interference benchmark assumes that the downlink (respectively, uplink) transmissions don't affect the IRS elements dedicated to assist the uplink (respectively, downlink) transmissions. This is idealistic as the uplink and downlink signals are reflected by all of the IRS elements and hence the phase shift associated with each element affects both uplink and downlink  transmissions. Conversely, the slicing-with-interference benchmark then computes  the achievable DL and UL rates using the entire IRS configuration. This in turn incorporates the interference terms due to slicing the IRS.}
 \section{Simulation results} \label{sec 5}
 \begin{figure*}%
\centering
\begin{subfigure}{0.99\columnwidth}
\captionsetup{justification=centering,margin=1cm}
\centerline{\includegraphics[width=\linewidth]{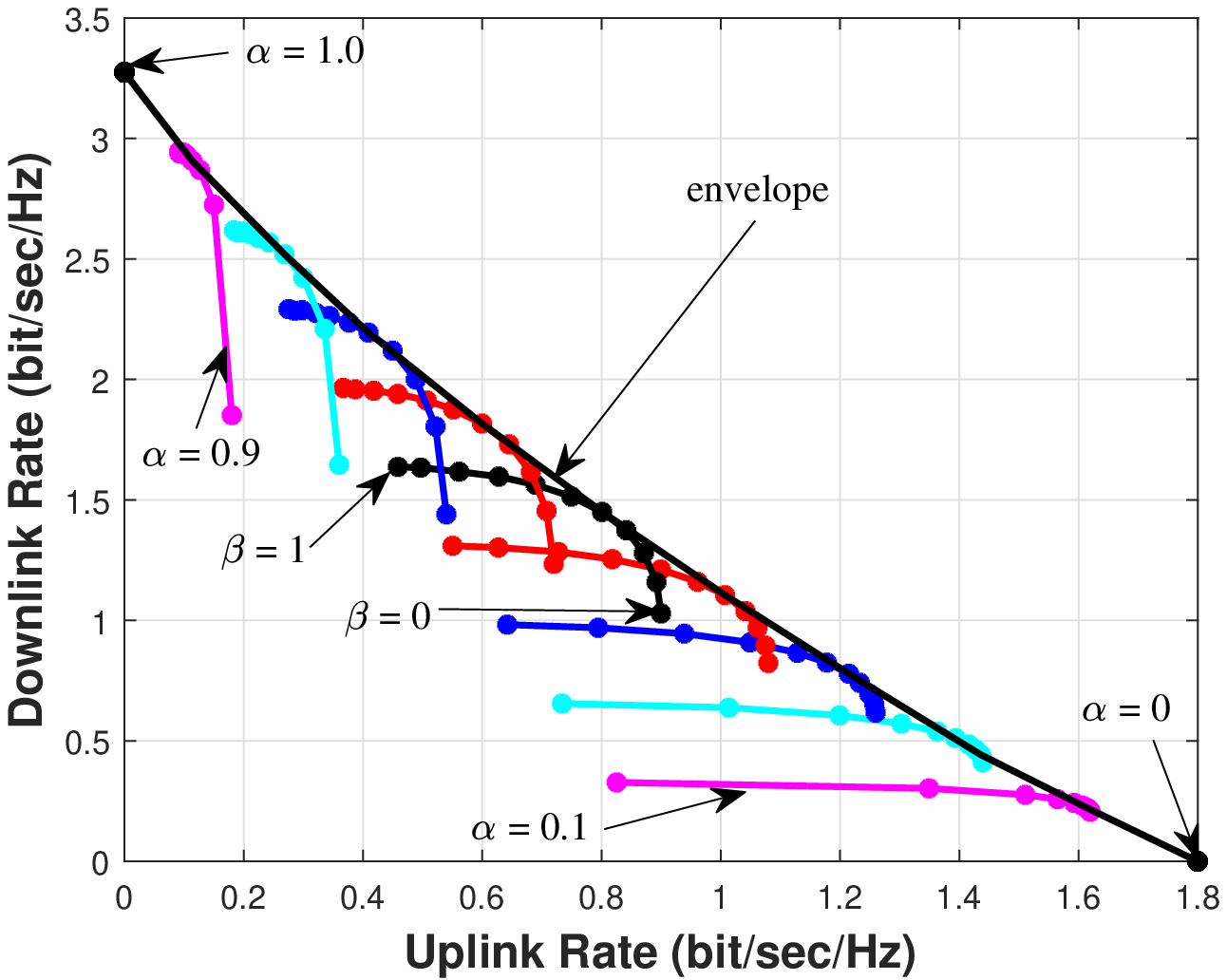}}
\subcaption{FDD}
\label{fig:Plot for alpha = 0.5}
\end{subfigure} \hfill 
\begin{subfigure}{0.99\columnwidth}
\captionsetup{justification=centering,margin=1cm}
\centerline{\includegraphics[width=\linewidth]{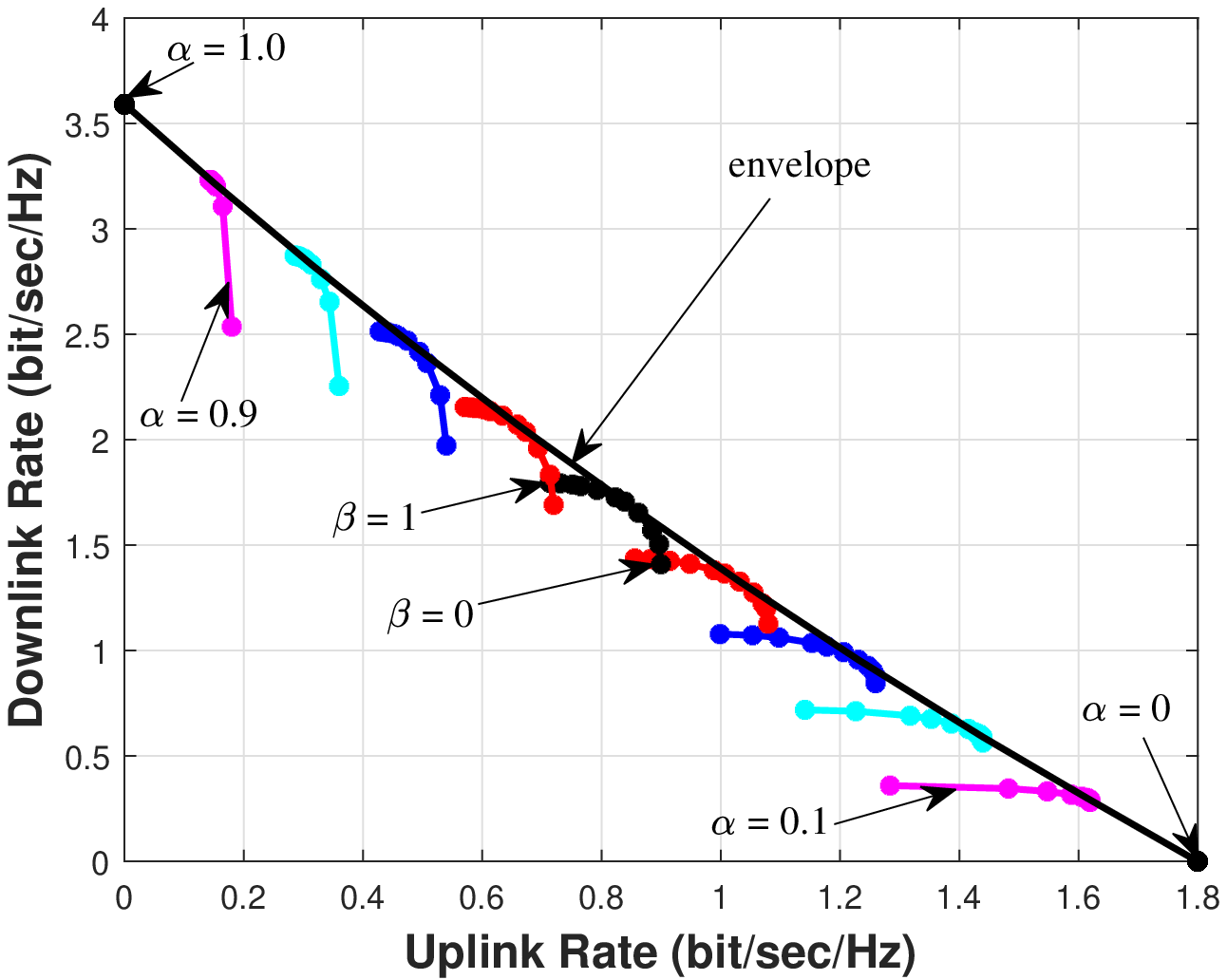}}
\subcaption{TDD}
\label{fig:Plot for alpha = 0.7}
\end{subfigure}
\caption{\textcolor{black}{Uplink-Downlink trade-off region for $\alpha = 0,0.1,\ldots,1$, where for a fixed $\alpha$, $\beta = 0,0.1,\ldots,1$, with $N=200$ and under independent weights.}}
\end{figure*}
\subsection{System setup and parameters} \label{subsec-4-A}
The IRS-assisted communications system is comprised of an $M$-antenna BS, an $N$-element IRS, and $K$ single-antenna users placed in a 3D coordinate system. Following Fig.~\ref{fig:System layout}, \textcolor{black}{the BS is located at $(x_{BS},y_{BS},h_{BS}) = (0,25, 25) \; m$ and the IRS is placed at $(x_{IRS}, y_{IRS}, h_{IRS}) =(300,0,15)\;m$. Moreover, the $K$ users are placed at a height of $h_c=1.5 \; m$ while being distributed within an $x-y$ plane disc which is centered at $(x_c,y_c)= (300,25) \; m$ with radius $R=20 \;m$}.
\par
A uniform rectangular array (URA) is used at the BS with $M$ elements such that $M_y \times M_z=M$ where  $M_y$ and $ M_z$ correspond to the number of elements aligned with the $y$-axis and $z$-axis. Similarly, a URA is used at the IRS with $N$ elements such that $N_x \times N_z=N$ where  $N_x$ and $ N_z$ correspond to the number of elements aligned with the $x$-axis and $z$-axis. 
\par
For TDD, the uplink/downlink carrier frequency is $1.95 $ GHz. In FDD, the uplink carrier frequency is $f_{UL}=1.95$ GHz and the downlink carrier frequency is $f_{DL}=2.14$ GHz. In both TDD and FDD, the antennas at the BS and the IRS are spaced by $ \frac{1}{2} \times \frac{c}{f_{UL}}$ where $c$ is the speed of light in free space. 
 \textcolor{black}{  Let $P_1 = 30$ dBm, $P_2 = 17$ dBm. In TDD, $P_{max}^{DL}= P_1$ and $P_{max}^{UL}=P_2$ and in FDD, {$P_{max}^{DL} =  \alpha P_1$  and $P_{max}^{UL} = (1-\alpha) P_2$ }. \textcolor{black}{Moreover, $N_0 = N_i \times NF$ where $N_i =-170$ dBm/Hz and the noise figure in downlink is $NF =9$ dB and in uplink $NF=7$ dB.}} \textcolor{black}{ For TDD, the bandwidth is $20$ MHz. In FDD, the total bandwidth is $20$ MHz, shared between uplink and downlink transmissions.} 
 \par 
 3GPP path loss models are used \cite{3gpp} with the line-of-sight (LOS) and non-line-of-sight (NLOS) models are given by
\begin{subequations}
    \begin{align}
    \label{eq:LOS_PL}
         PL^{LOS}(d_{3d},f_{c}) &= 28 + 22\log_{10}(d_{3D})+20 \log_{10}(f_c), \qquad \\
    \label{eq:NLOS_PL}
       PL^{NLOS}(d_{3d},f_{c}) &= 13.54 + 39.08 \log_{10}(d_{3D}) \nonumber \\
        & \quad + 20  \log_{10}(f_c) - 0.6(h_{c}-1.5)
        \end{align}
\end{subequations} 
where $d_{3d}$ is the 3D distance between two points and $f_c$ is the carrier frequency.
For each user $k \in \mathcal{K}$, the IRS-user reflected channel $\boldsymbol{h}_{t,r,k}$ and the BS-IRS channel $\boldsymbol{G}_{t}$ are modeled using a Rician fading model for $t \in \{UL,DL\}$. In particular,
\begin{subequations}
    \begin{align}
      \boldsymbol{G}_{t} &=  \sqrt{PL^{LOS}(d_{BS-IRS},f_{t})} \nonumber \\
        & \quad \times \left(  \sqrt{\frac{\kappa}{1+\kappa}}  \boldsymbol{G}^{LOS}_{t}  + \sqrt{\frac{1}{1+\kappa}}  \boldsymbol{G}_{t}^{NLOS}  \right),   \\
       \boldsymbol{h}_{t,r,k} &=  \sqrt{PL^{LOS}(d_{{IRS-k}},f_{t})} \nonumber \\
        & \quad  \times \left( \sqrt{\frac{\kappa_k}{1+\kappa_k}}  \boldsymbol{h}_{t,r,k}^{LOS}   + \sqrt{\frac{1}{1+\kappa_k}}  \boldsymbol{h}_{t,r,k}^{NLOS}  \right), 
    \end{align}
\end{subequations}
where $P_{LOS}(d_{3d},f_{c})$ is given in \eqref{eq:LOS_PL}. Also, $d_{BS-IRS}$ denotes the 3D distance between the BS and the IRS and $d_{BS-k}$ denotes the 3D distance between the IRS and user $k$. The Rician factors $\kappa$ and $\{\kappa_k\}_{k=1}^K$ capture the relative strength of the LOS component compared to the NLOS components between the BS and the IRS and between the IRS and user $k$, respectively. The matrix $\boldsymbol{G}^{LOS}_{t} \in \mathbb{C}^{M \times N}$ contains the phases for the LOS components which were computed using the underlying geometry. \textcolor{black}{In particular, for the $m^{th}$ element of the BS and the $n^{th}$ element of the IRS, separated by 3D distance of $d_{3d}^{\; m,n}$, the corresponding phase shift is given by $    \left [\boldsymbol{G}^{t}_{LOS}  \right ]_{m,n} = \exp \left(  \frac{ 2 \; \pi \; j \; d_{3d}^{ \; m,n}}{\lambda_t} \right)$
where $\lambda_t$ is the wavelength associated with the corresponding carrier frequency.} Similarly, the vector $\boldsymbol{h}_{t,r,k}^{LOS}$ captures the $N$ phase shifts associated with LOS component between user $k$ and the IRS. On the other hand, $\boldsymbol{G}_{t}^{NLOS} $ and $\boldsymbol{h}_{t,r,k}^{NLOS}$ capture the NLOS components associated with BS-IRS and IRS-user $k$ propagation channels, respectively. Moreover, the Rayleigh fading components $\boldsymbol{G}_{t}^{NLOS} $ and $\boldsymbol{h}_{t,r,k}^{NLOS}$ are sampled with each entry i.i.d from $\mathcal{CN}(0,1)$. 

\begin{figure*}%
\centering
\begin{subfigure}{.99\columnwidth}
\captionsetup{justification=centering,margin=1cm}
\centerline{\includegraphics[width=\linewidth]{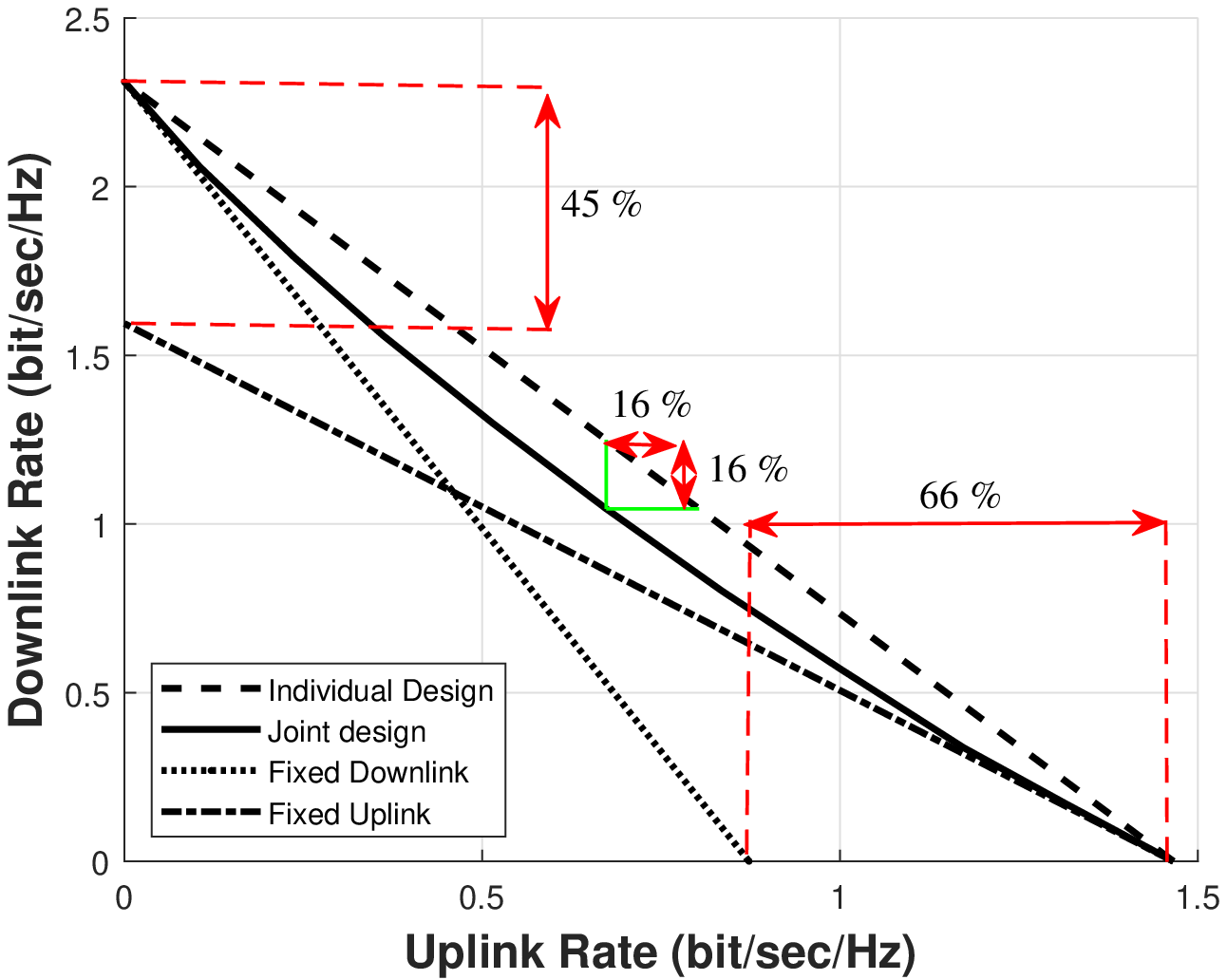}}
\subcaption{FDD, $N=200$ }
\label{fig:Plot for N=200 FDD equal}
\end{subfigure} \hfill 
\begin{subfigure}{.99\columnwidth}
\captionsetup{justification=centering,margin=1cm}
\centerline{\includegraphics[width=\linewidth]{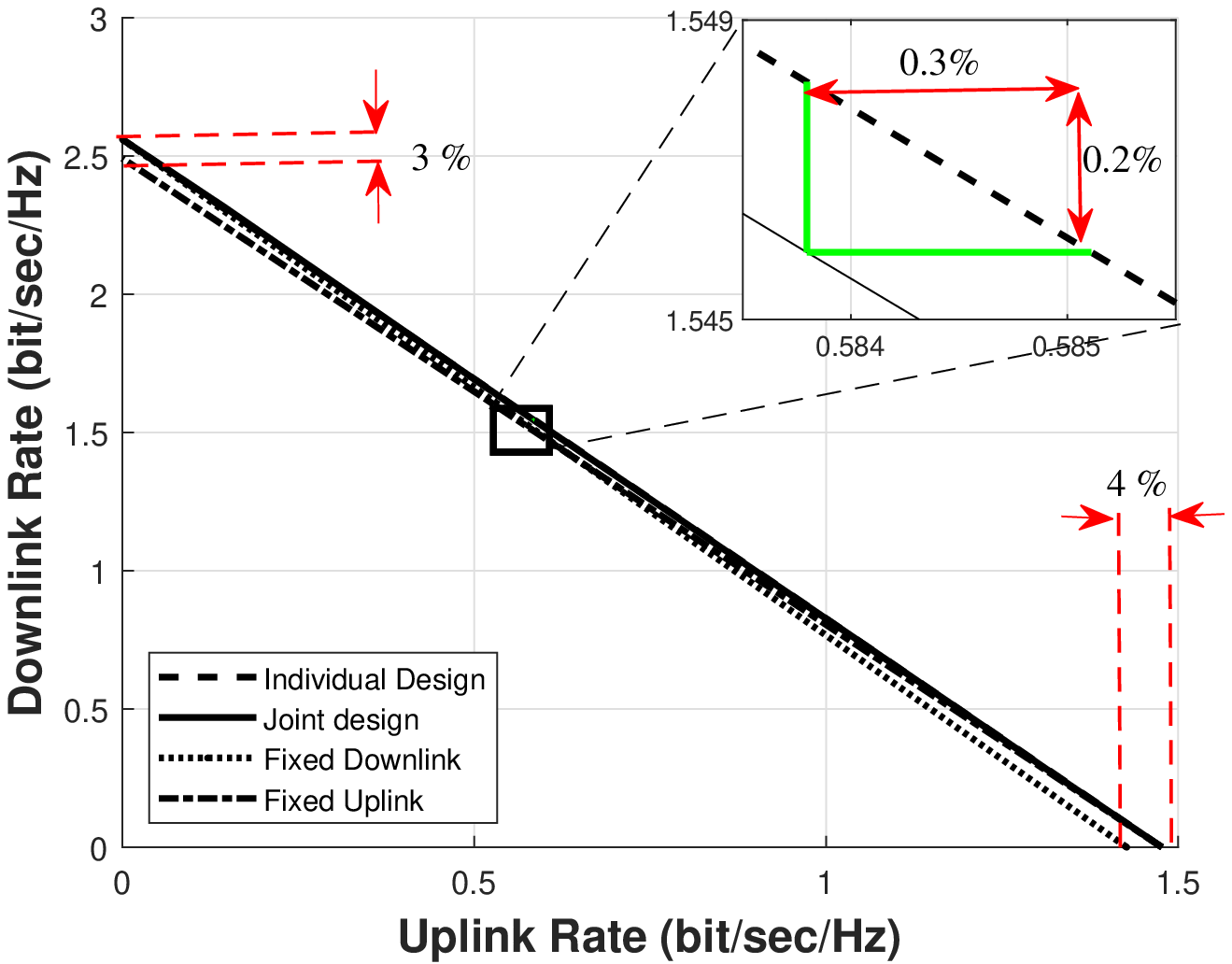}}
\subcaption{TDD, $N=200$ }
\label{fig:Plot for N=200 TDD equal}
\end{subfigure} 
\caption{Uplink-Downlink Rate region for  $N=200$ and equal weights.}
\label{equal plots}
\end{figure*}

\begin{figure*}%
\centering
\begin{subfigure}{.99\columnwidth}
\captionsetup{justification=centering,margin=1cm}
\centerline{\includegraphics[width=\linewidth]{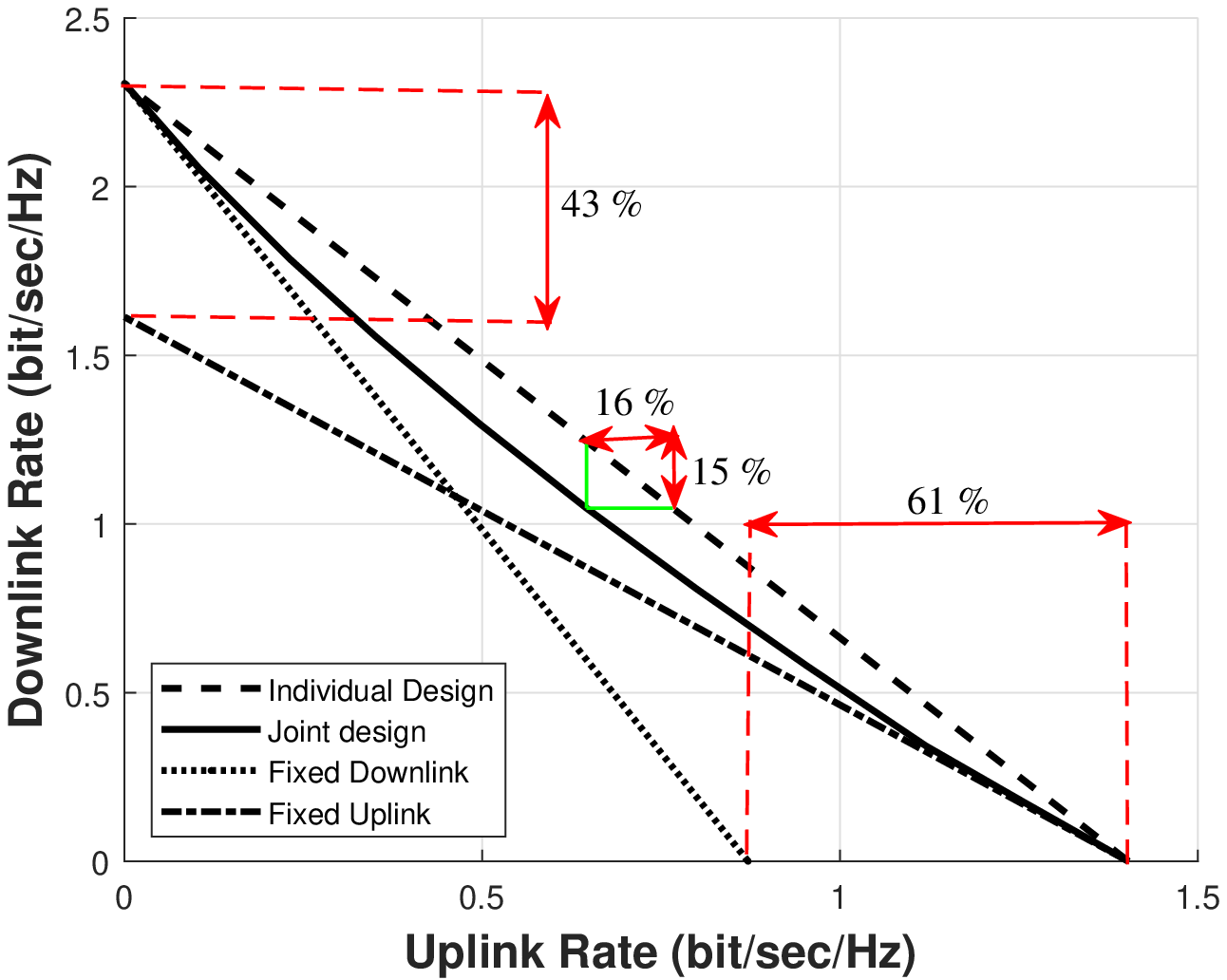}}
\subcaption{FDD, $N=200$ }
 \label{fig:Plot for N=200 PF}
\end{subfigure} \hfill 
\begin{subfigure}{.99\columnwidth}
\captionsetup{justification=centering,margin=1cm}
\centerline{\includegraphics[width=\linewidth]{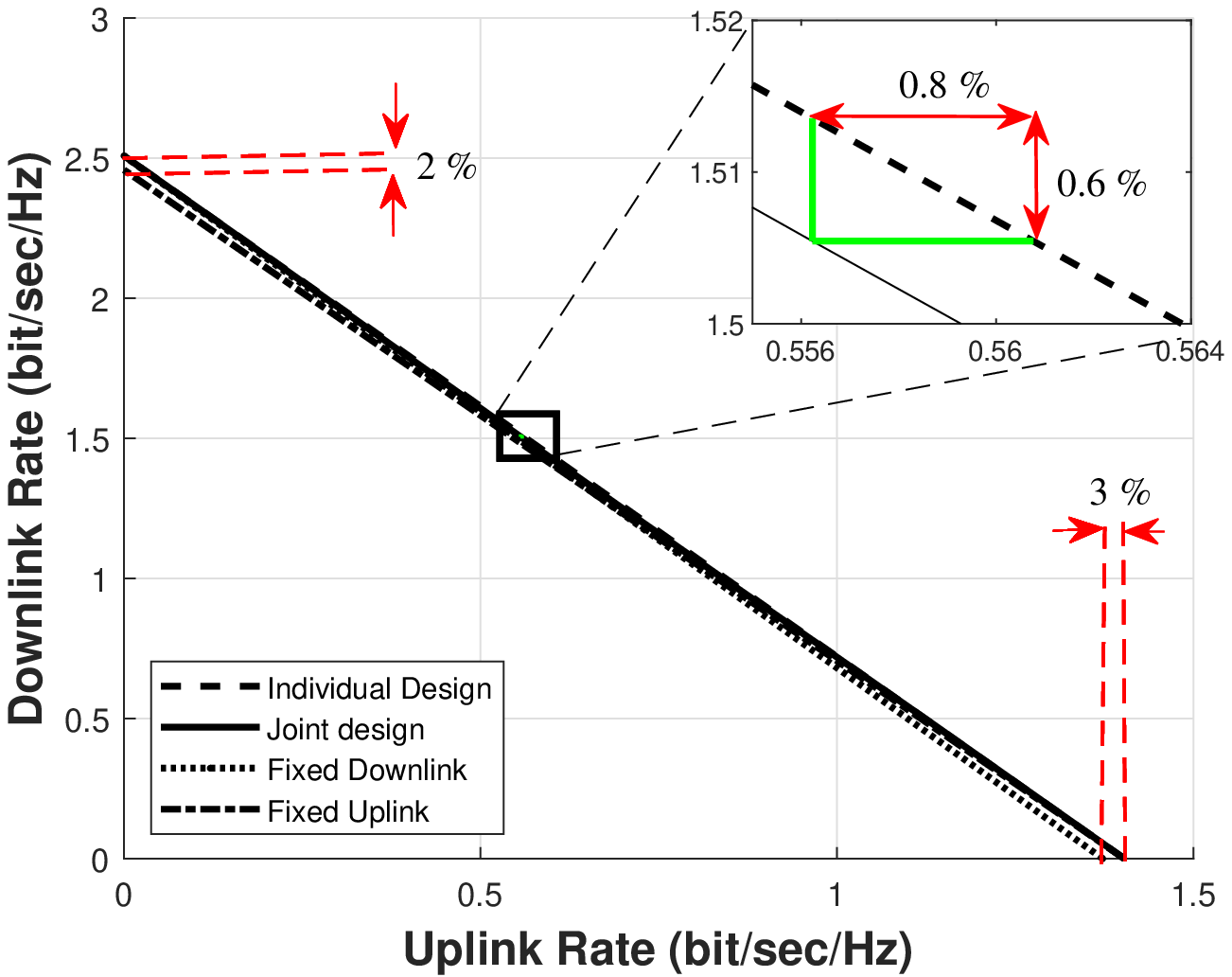}}
\subcaption{TDD, $N=200$ }
\label{fig:Plot for N=200 PF TDD }
\end{subfigure}
\caption{Uplink-Downlink Rate region for $N=200$ and PF weights.}
\label{PF plots}
\end{figure*}
\par 
For each user $k$, the BS-user channel $\boldsymbol{h}_{t,d,k}$ is modeled using Rayleigh fading. Particularly, 
\begin{equation}
    \begin{aligned}
    \boldsymbol{h}_{t,d,k} = \sqrt{PL^{NLOS}}   \;  \boldsymbol{h}_{t,d,k}^{NLOS},
    \end{aligned}
\end{equation}
where $PL^{NLOS}$ is given in \eqref{eq:NLOS_PL} and the components of  $\boldsymbol{h}_{t,d,k}^{NLOS}$ are sampled from $\mathcal{CN}(0,1)$. \textcolor{black}{ Unless otherwise specified, the Rician factors are $\kappa = 6\;$dB and $\{\kappa_k\}_{k=1}^{K} = 8\;$dB, and the number of antennas at the BS and IRS are $M= 4 \times 2=8$ and $N= 20 \times 10 =200$ serving $K=4$ users. \textcolor{black}{The numerical results were obtained by averaging the results over 100 independent channel realizations.}}

\subsection{Uplink-Downlink Trade-off Regions}
\label{Trade-off region Analysis}
\par
For a fixed time/frequency resource allocation defined by a fixed $\alpha$, the relative priority of uplink vs downlink is defined by the weight $\beta$. In order to first explore the rate region for a fixed $\alpha$, the problem in \eqref{eq:UL_DL_WSR} is {optimized} for different values of $\beta$ from 0 to 1 under the {independent} weighting strategy. Figs.~\ref{fig:Plot for alpha = 0.5} and \ref{fig:Plot for alpha = 0.7} \textcolor{black}{show the trade-off between uplink weighted rates, given by (10a) and (8b), versus downlink weighted rates, given by (9a) and (8a), for $\alpha= 0,0.1,\ldots,1$ and both FDD and TDD systems.}  
\begin{figure*}%
\centering
\begin{subfigure}{.99\columnwidth}
 \captionsetup{justification=centering,margin=1cm}
\centerline{\includegraphics[width=1\linewidth]{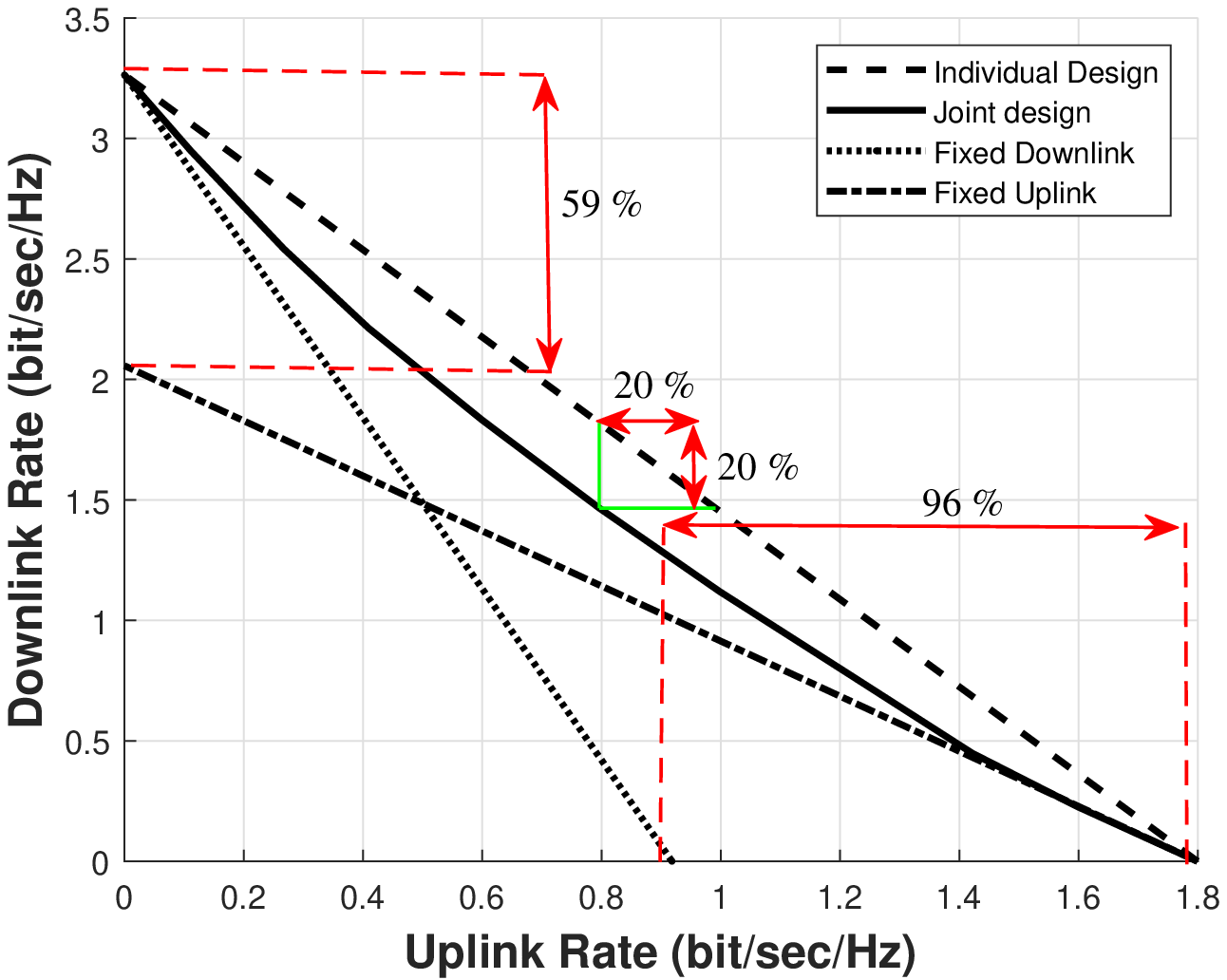}}
\subcaption{FDD, $N=200$ }
\label{FDD fig:Plot for N=200 disp}
\end{subfigure}\hfill%
\begin{subfigure}{.99\columnwidth}
 \captionsetup{justification=centering,margin=1cm}
\centerline{\includegraphics[width=\linewidth]{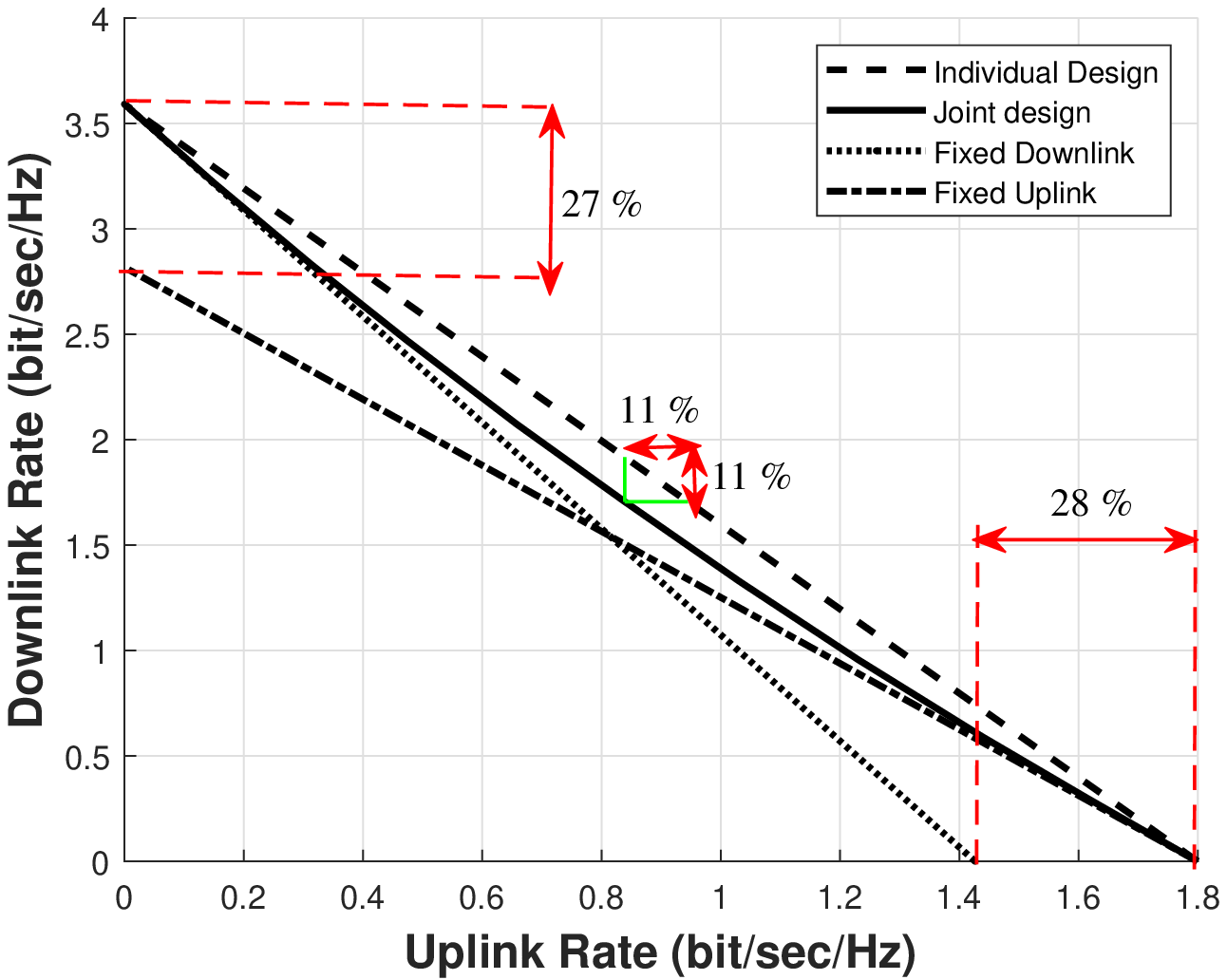}}
\subcaption{TDD, $N=200$ }
\label{fig:TDD Plot for N=200 disp}
\end{subfigure}
\begin{subfigure}{.99\columnwidth}
\captionsetup{justification=centering,margin=1cm}
\centerline{\includegraphics[width=1\linewidth]{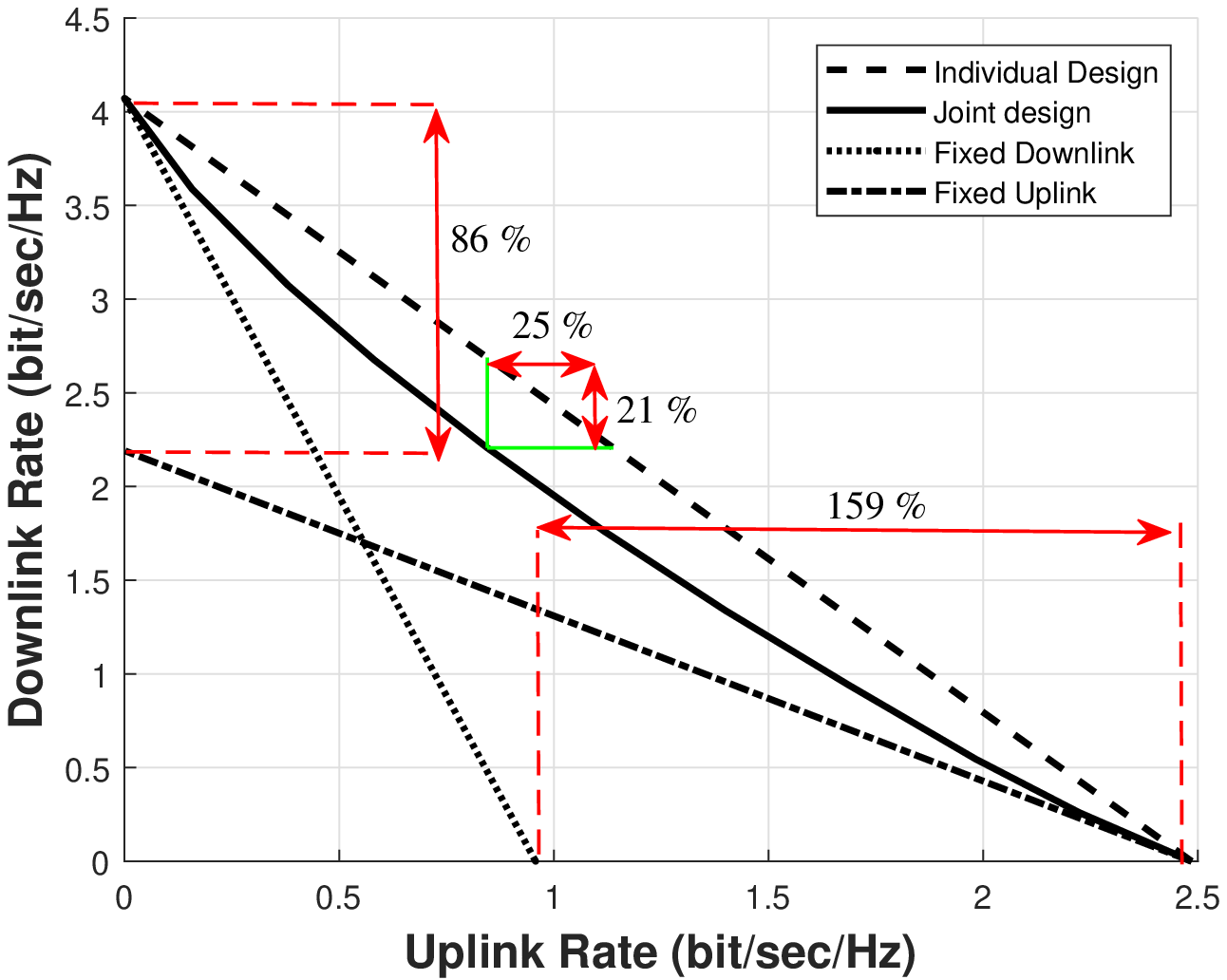}}
\subcaption{FDD, $N=400$ }
\label{FDD fig:Plot for N=400 disp}
\end{subfigure}\hfill%
\begin{subfigure}{.99\columnwidth}
     \captionsetup{justification=centering,margin=1cm}
    \centerline{\includegraphics[width=\linewidth]{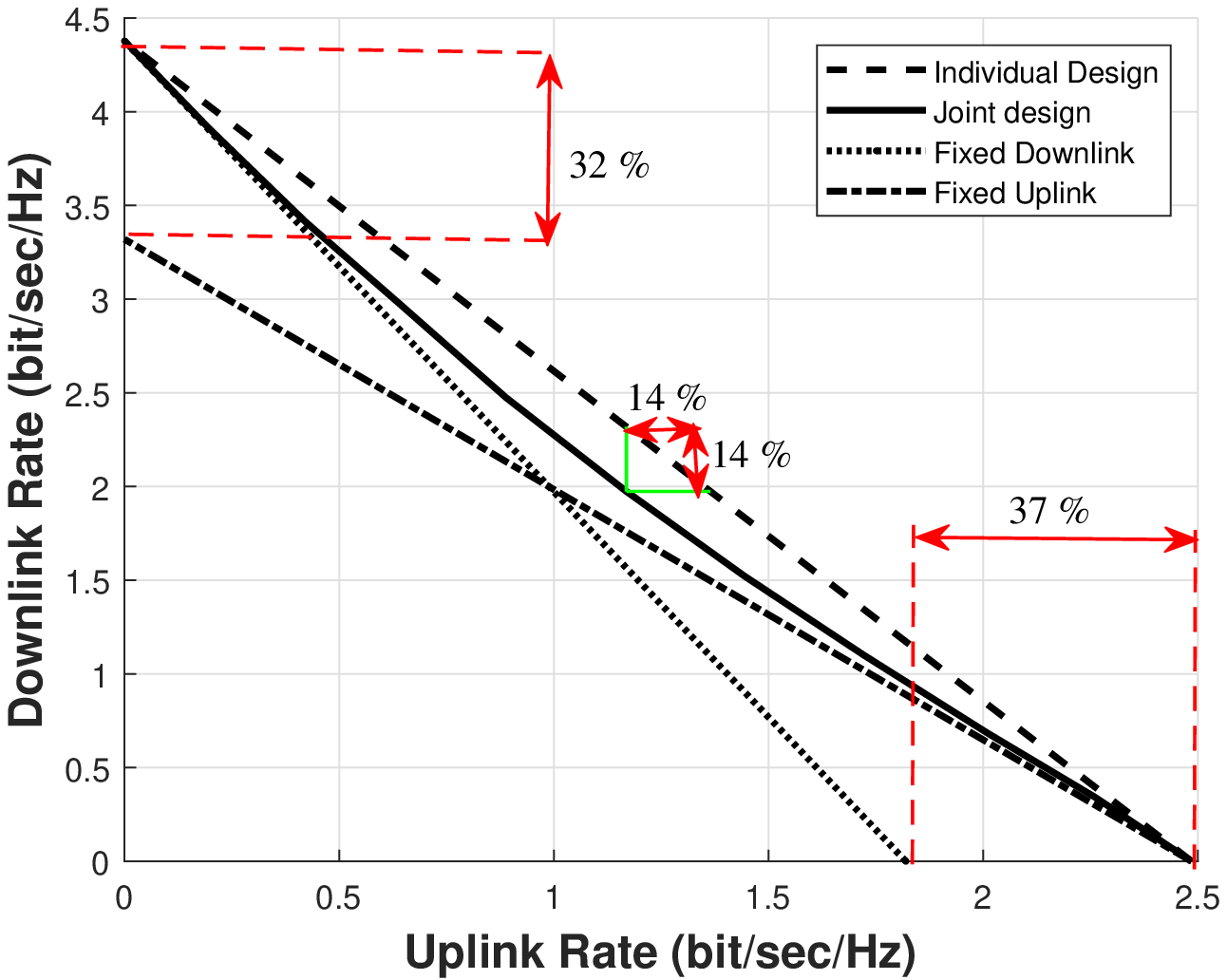}}
    \subcaption{TDD, $N=400$ }
    \label{fig:TDD Plot for N=100 disp}
\end{subfigure}
\caption{Uplink-Downlink region for $N=200$ and $N=400$, under independent weighting.}
\label{independent plots}
\end{figure*}

\par
\textcolor{black}{
By optimizing the fraction  of  time or band  spent  on uplink  vs downlink, i.e. optimizing $\alpha$, the entire trade-off between UL-rate and DL-rate {achieved by the BCD algorithm} is found} and is given by the envelope of the curves in Figs.~\ref{fig:Plot for alpha = 0.5} and \ref{fig:Plot for alpha = 0.7}, and shown in solid black. {This envelope represents the trade-off between UL weighted rates and DL weighted rates under a joint design and UL/DL resource optimization.}
\textcolor{black}{The performance of the joint design envelope with respect to the individual design and the fixed-downlink/fixed-uplink schemes are shown in Figs.~\ref{equal plots}-\ref{independent plots}}
\textcolor{black}{for the three user weighting schemes.} \textcolor{black}{Four metrics are used to compare performance: the maximum DL gain of joint design over fixed-uplink, the maximum UL gain of joint design over fixed-downlink, and both the maximum DL loss and the maximum UL loss} of joint design compared to individual design. The two maximum gain metrics correspond to the maximum improvement yielded by employing a joint design vs employing a {fixed-downlink/fixed-uplink} design. \textcolor{black}{Moreover, the loss of performance due to joint design compared with the {individual design} is quantified using the maximum DL/UL losses.}

\par
Figs.~\ref{fig:Plot for N=200 FDD equal} and \ref{fig:Plot for N=200 TDD equal}  show {{the} trade-off curves} with equal weighting strategies for FDD and TDD systems with $N=200$ along with the 4 performance metrics described above. \textcolor{black}{Figs.~\ref{fig:Plot for N=200 PF} and \ref{fig:Plot for N=200 PF TDD } show {{the} trade-off curves} associated with PF weighting under FDD and TDD systems, where the PF weights were obtained using the approximation in Section \ref{sec: user weighting} for a large number of time slots $s=100$. Similarly, Figs.~\ref{FDD fig:Plot for N=200 disp} and \ref{fig:TDD Plot for N=200 disp} show {{the} trade-off curves} with independent weighting, again for $N=200$.} {From the figures, }\textcolor{black}{ the improvement due to joint design compared with the fixed-downlink/uplink designs is more significant as \textcolor{black}{we deviate towards independent weights and as more flexibility in resource allocation is needed.}} \textcolor{black}{{Indeed,} as shown in Fig.~\ref{independent plots}, the maximum UL gain due to joint design is {$96 \%$ and the maximum DL gain is $59\%$ when independent weights are used under the FDD scenario.} This is compared to $61 \%$ and $43\%$ when PF weights are used (Fig.~\ref{fig:Plot for N=200 PF}) and compared to $66 \%$ and $45\%$ when equal weights are used (Fig.~\ref{fig:Plot for N=200 FDD equal}). The same pattern applies for TDD where the maximum UL and DL gains are $28 \%$ and $27\%$ when independent weights are used. This is in contrast with max-UL and max-DL gains of $3 \%$ and $2\%$ when PF weights are used and $4 \%$ and $3\%$ when equal weight are used as shown in Figs.~\ref{fig:Plot for N=200 PF TDD } and \ref{fig:Plot for N=200 TDD equal}, respectively}.

\par
\textcolor{black}{{As outlined in Section \ref{3C}, TDD benefits from channel reciprocity and hence the max-UL and max-DL losses, of joint over individual design, are  expected to be lower for TDD compared to FDD.} Under TDD, the max-UL loss is $0.3 \%$, $0.8 \%$, and $11 \%$ for equal, PF, and independent weights, respectively. This is compared to $16 \%$, $16 \%$, and $20 \%$ for FDD. Similarly, for TDD  the max-DL loss is $0.2 \%$, $0.6 \%$ and $11 \%$ for equal, PF, and independent weights, respectively. This is in contrast with the higher losses in FDD of $16 \%$, $15 \%$, and $20 \%$.}    
\par 
As shown in Figs.~\ref{FDD fig:Plot for N=400 disp} and \ref{fig:TDD Plot for N=100 disp} compared to Figs.~\ref{FDD fig:Plot for N=200 disp} and \ref{fig:TDD Plot for N=200 disp}, the joint design provides greater benefits for larger $N$. \textcolor{black}{ The max-DL and max-UL gains increase from $59\%$ and $96\%$ for $N=200$ to $86\%$ and $159\% $ for $N=400$ in the case of FDD configuration.} \textcolor{black}{We attribute this to the fact that as $N$ increases, performance is more dependent on the IRS. Hence, a  joint IRS design with optimized phase shifts becomes more beneficial compared to fixed-downlink/fixed-uplink designs.}
\begin{figure*}%
\centering
\label{FDD_overN}
\begin{subfigure}{.99\columnwidth}
\captionsetup{justification=centering,margin=1cm}
\includegraphics[width=\linewidth]{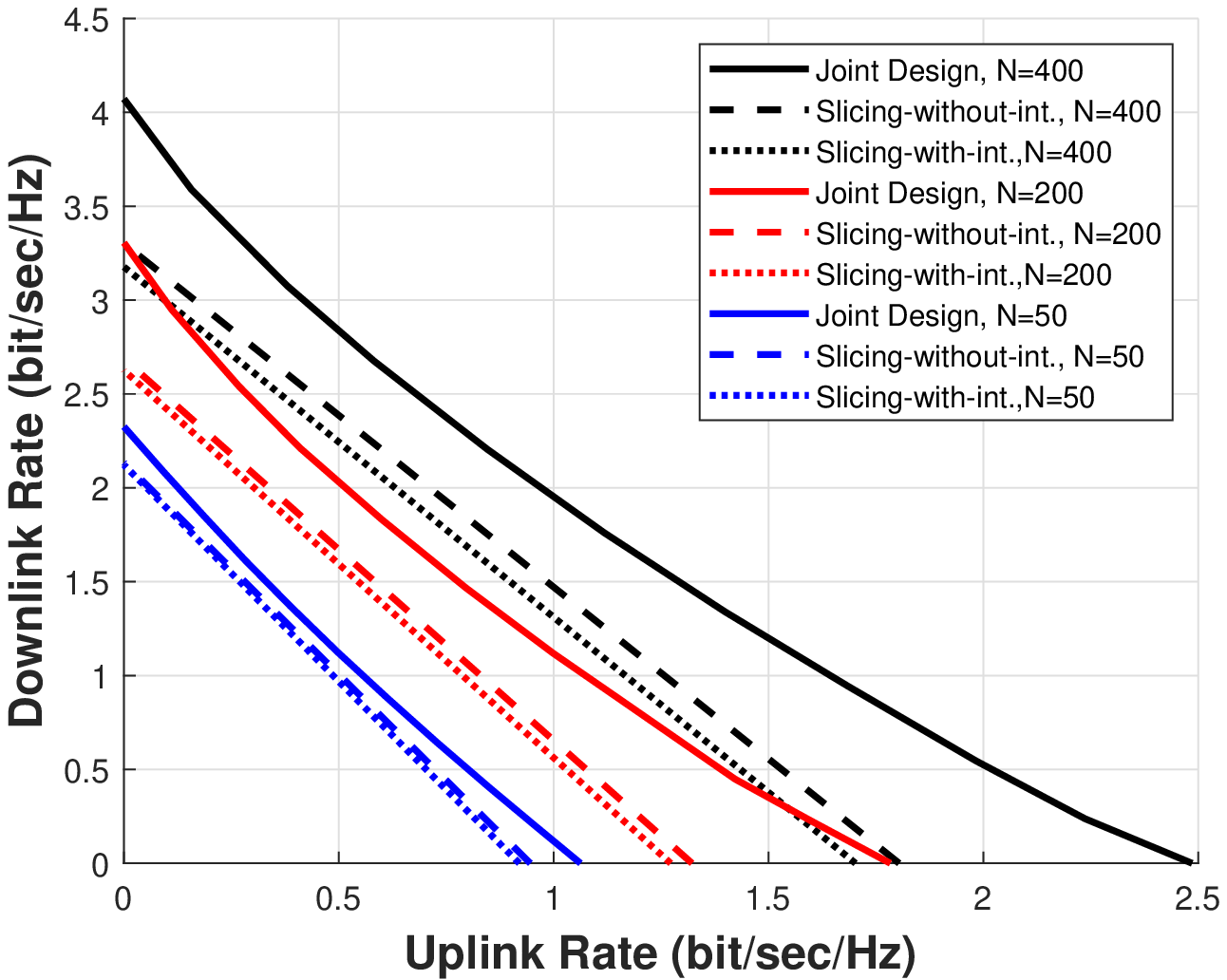}
\subcaption{ FDD }
\label{fig:Plot for  diff N }
\end{subfigure} \hfill 
\begin{subfigure}{.99\columnwidth}
\captionsetup{justification=centering,margin=1cm}
\includegraphics[width= \linewidth]{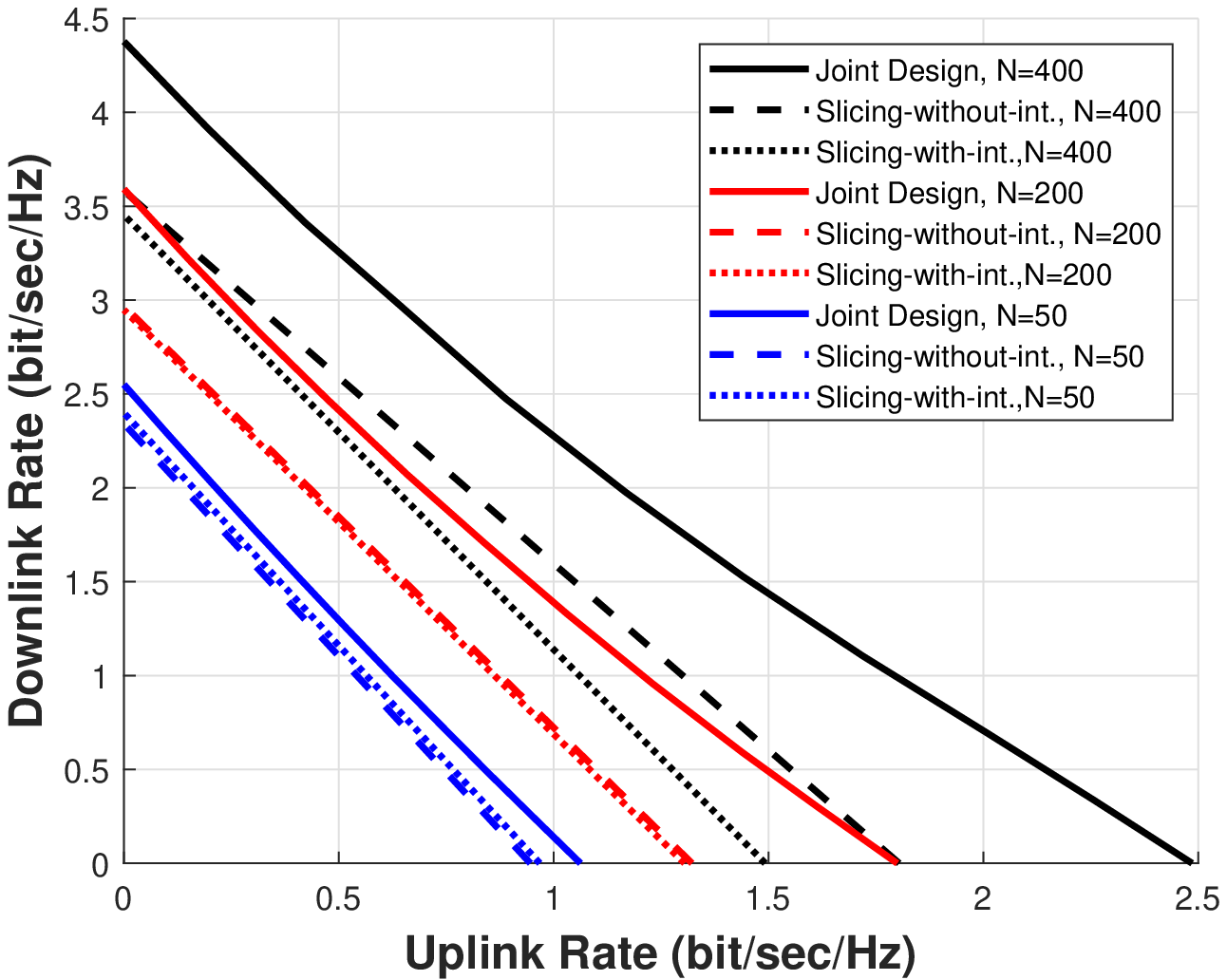}
\subcaption{ TDD }
\label{fig:Plot for diff N TDD}
\end{subfigure}
\caption{{Uplink-Downlink Rate region achieved by joint design and slicing benchmarks for different $N$ and independent weights.}}
 \label{Rate region for N}
\end{figure*}
\par
 \textcolor{black}{Fig.~\ref{Rate region for N} shows, for independent weights, {the uplink-downlink trade-off regions achieved by an optimized joint design and the two slicing benchmarks (slicing-with-interference and slicing-without-interference) for different values of $N$ under both FDD and TDD.} The figure reveals that for both FDD and TDD, the joint design outperforms the slicing-with-interference and the slicing-without-interference benchmarks. This follows since for all values of $N$ in Fig.~\ref{Rate region for N}, the UL-DL region  achieved by the joint design  strictly contains the UL-DL regions associated with the two slicing benchmarks. Furthermore, since slicing-without-interference underperforms joint design, Fig.~\ref{Rate region for N} also demonstrates that the interference alone is not responsible for the inferior performance achieved by a sliced IRS.}
 \begin{figure*}%
\centering
\begin{subfigure}{\columnwidth}
\captionsetup{justification=centering,margin=1cm}
\includegraphics[width=\linewidth]{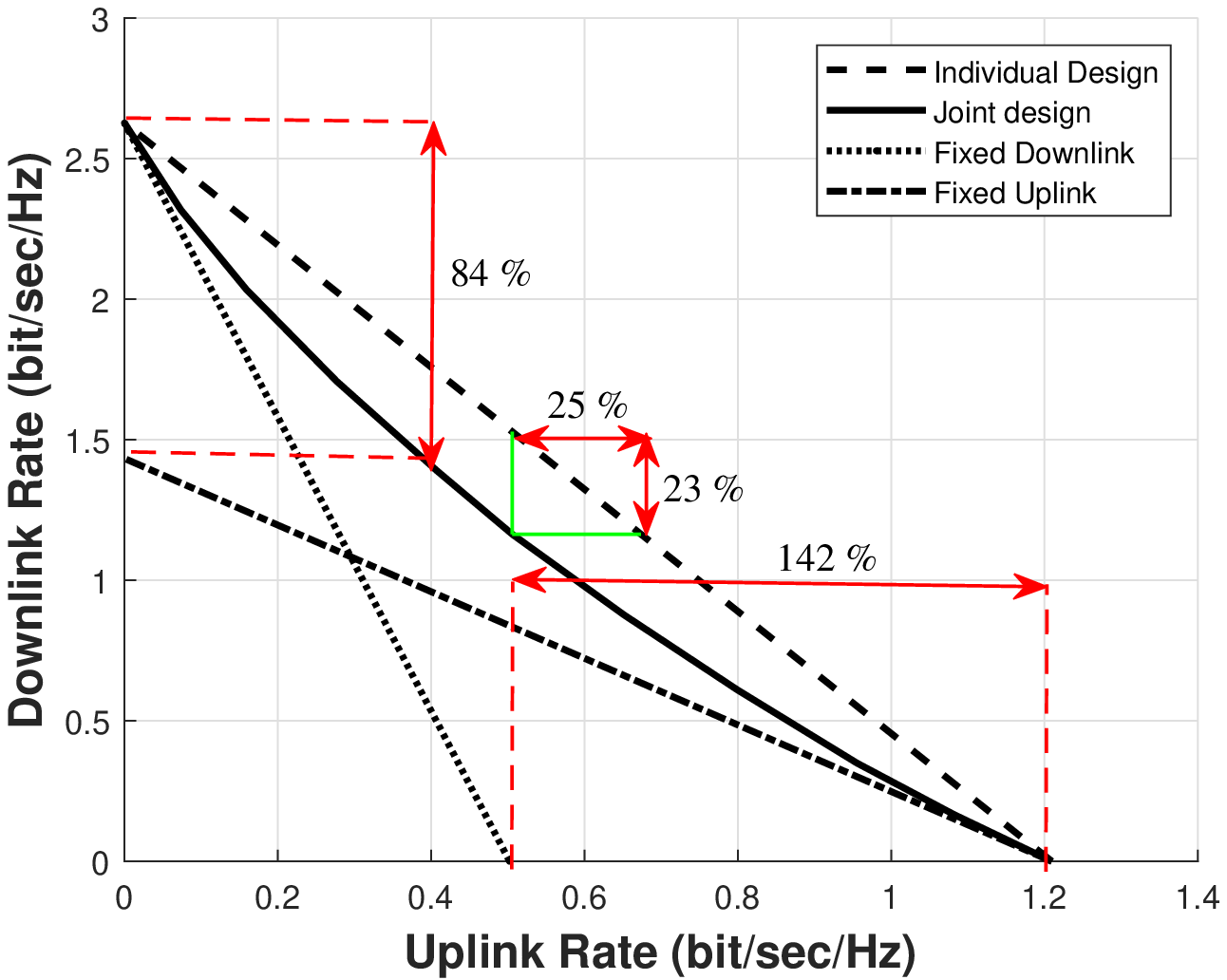}
\subcaption{ FDD, $M=4$ }
\label{11fig:Plot for N=200 M=4}
\end{subfigure}\hfill%
\begin{subfigure}{\columnwidth}
\captionsetup{justification=centering,margin=1cm}
 \includegraphics[width=\linewidth]{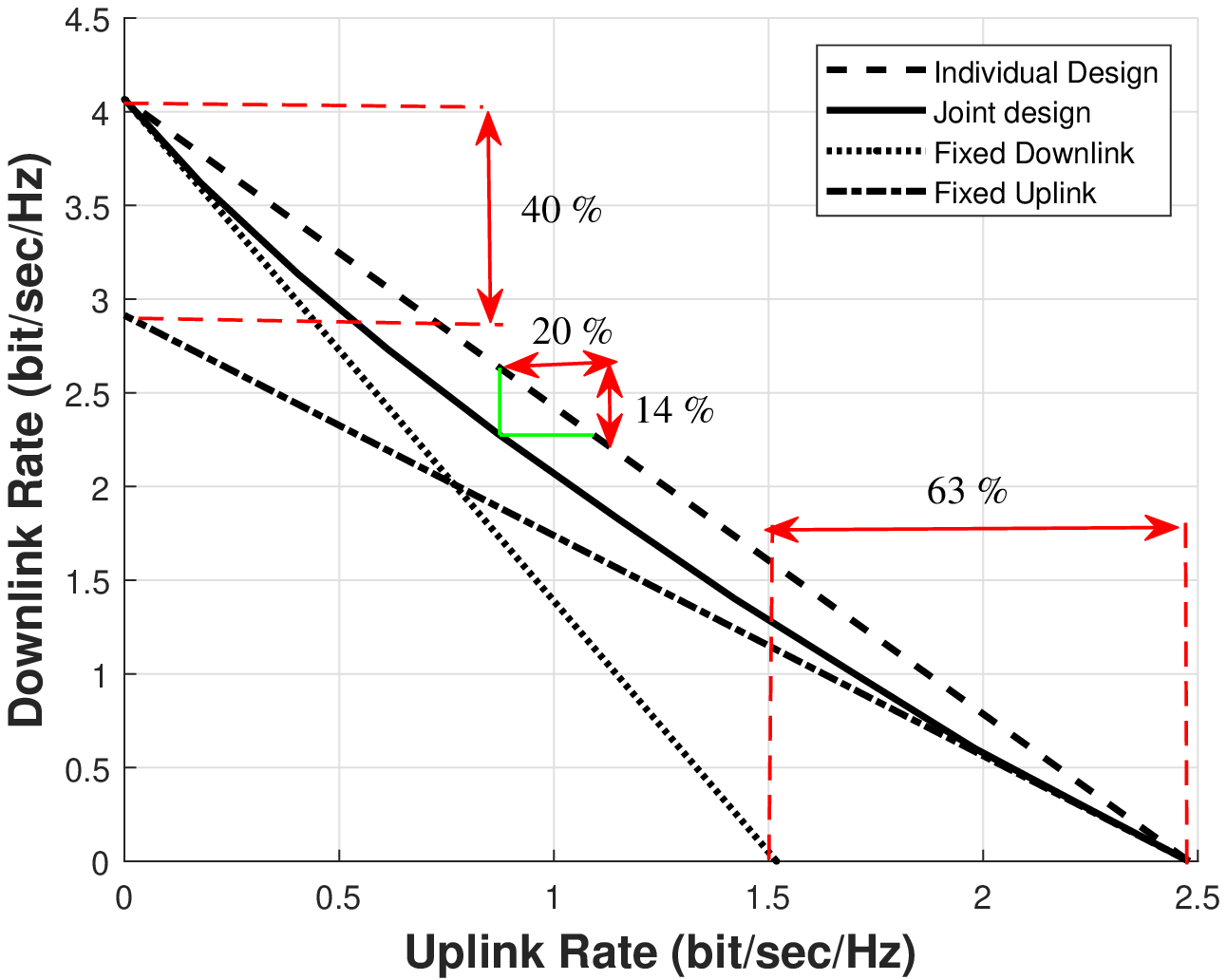}
 \subcaption{ FDD, $M=16$ }
 \label{11fig:Plot for N=200 M=16}
\end{subfigure}\hfill%
\begin{subfigure}{\columnwidth}
 \captionsetup{justification=centering,margin=1cm}
 \includegraphics[width=\linewidth]{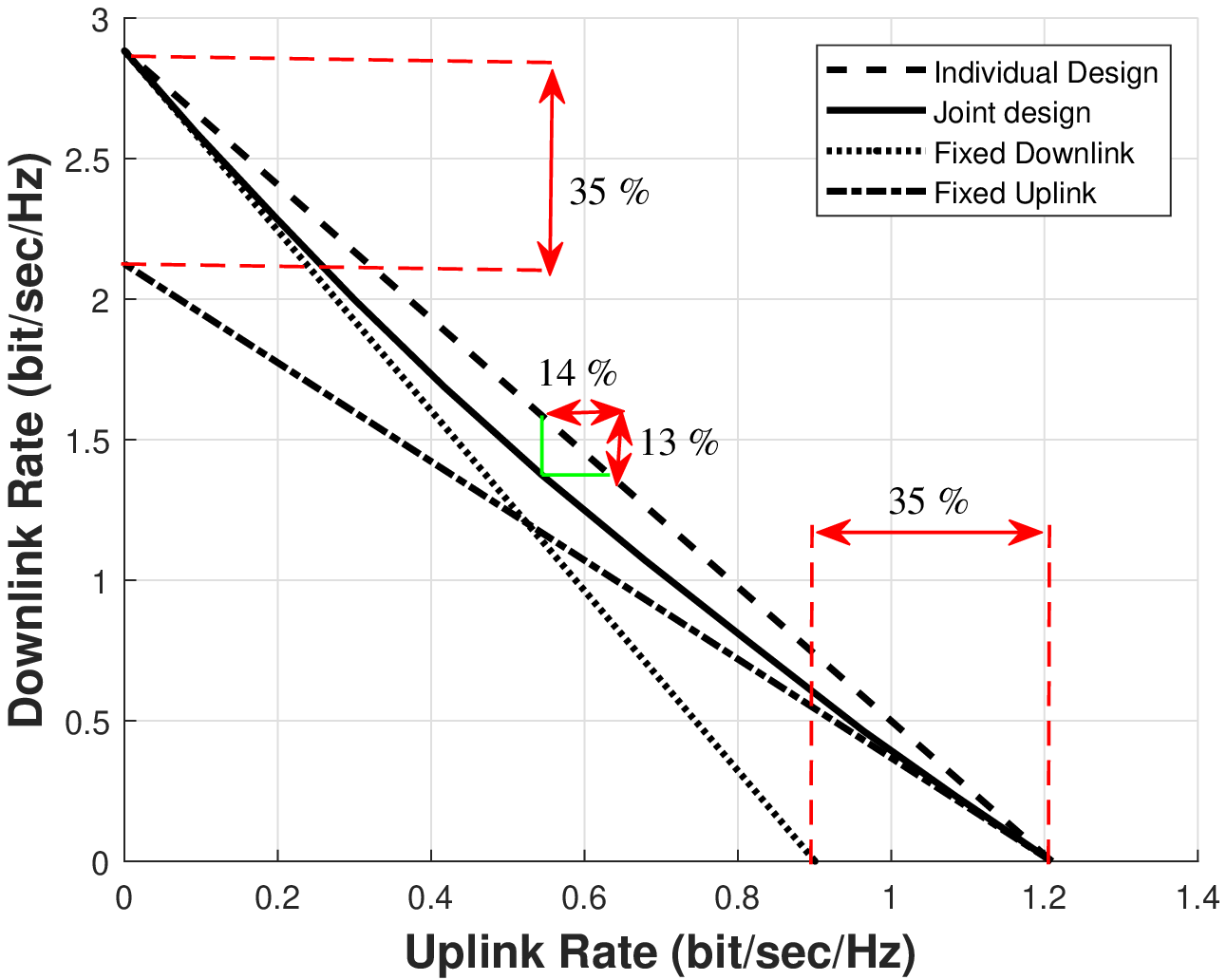}
 \subcaption{ TDD, $M=4$ }
 \label{TDD fig:Plot for N=200 M=4}
\end{subfigure}\hfill 
\begin{subfigure}{\columnwidth}
   \captionsetup{justification=centering,margin=1cm}
 \includegraphics[width=\linewidth]{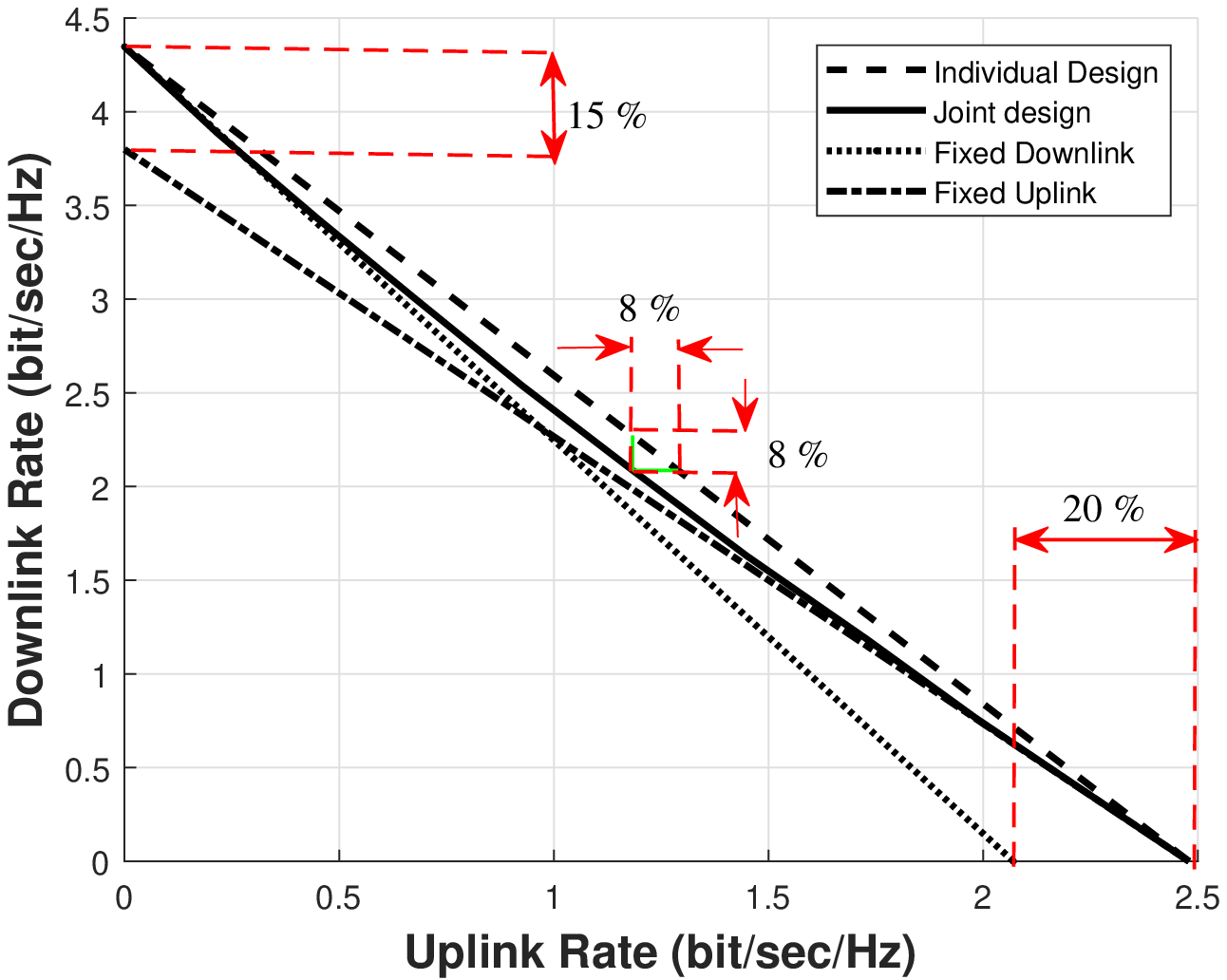}
 \subcaption{ TDD, $M=16$ }
 \label{TDD fig:Plot for N=200 M=16}
\end{subfigure} \hfill 
\caption{Uplink-Downlink Trade-off region for $N=200$ under FDD/TDD for different $M$ with independent weights.}
\label{fig: Different M}
\end{figure*}
\begin{figure*}%
\centering
\centering
\label{ZF}
\begin{subfigure}{.99\columnwidth}
\captionsetup{justification=centering,margin=1cm}
\includegraphics[width=\linewidth]{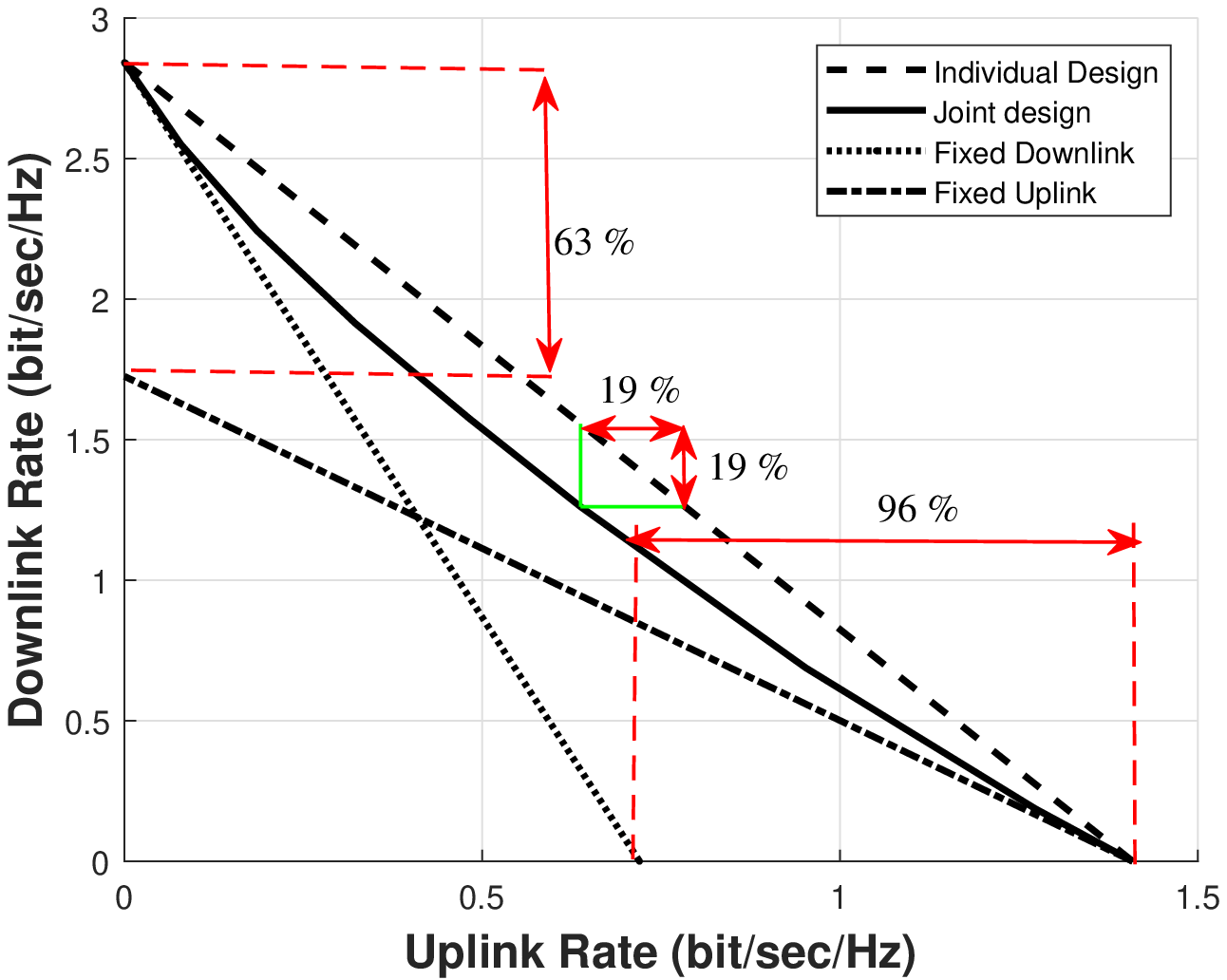}
\subcaption{ FDD }
\label{fig:Plot for N=200 ZF}
\end{subfigure} \hfill 
\begin{subfigure}{.99\columnwidth}
\captionsetup{justification=centering,margin=1cm}
\includegraphics[width= \linewidth]{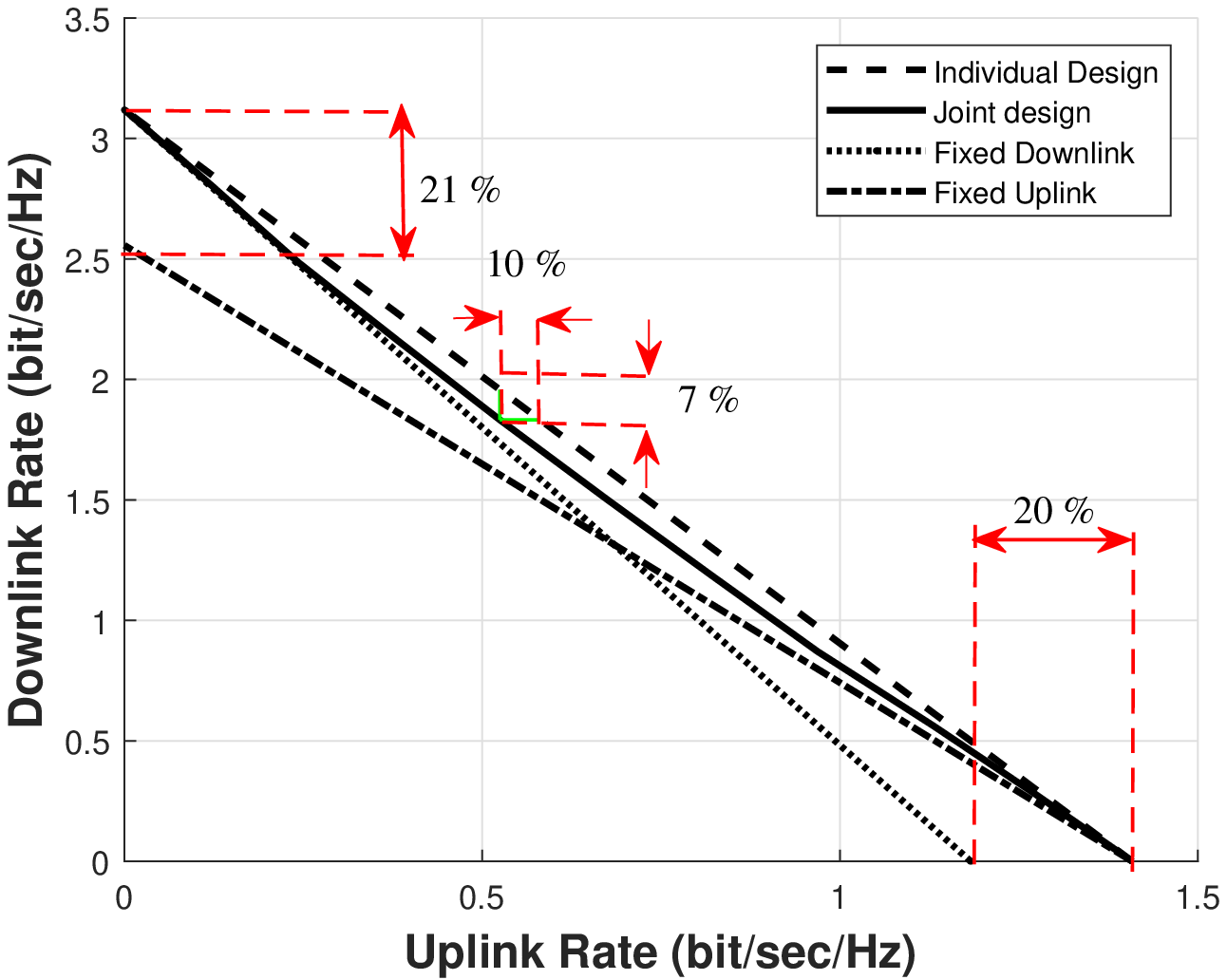}
\subcaption{ TDD }
\label{fig:Plot for N=200 ZF TDD}
\end{subfigure}
\caption{Uplink-Downlink Rate region for $N=200$ under ZF beamforming and independent weighting. }
\label{Fig: COmpare ZF}
\end{figure*}

\par
\textcolor{black}{ Fig.~\ref{fig: Different M} shows the impact of the number of antennas at the BS \textcolor{black}{on the max-UL and max-DL gains of joint design over fixed-uplink/fixed-downlink designs as well as the max-UL and max-DL losses of joint design over the individual design, under independent weights. From the figure, as the ratio of passive beamforming elements to active beamforming elements is increased, the performance of the system becomes more dependent on the passive beamforming at the IRS and hence the performance improvement due to joint IRS design is more substantial.} Specifically, Figs.~\ref{11fig:Plot for N=200 M=4} and \ref{11fig:Plot for N=200 M=16} show the uplink-downlink trade-off regions for $M=4$ and $M=16$ under FDD with an IRS of $N=200$ elements. {Comparing Figs.~\ref{11fig:Plot for N=200 M=4} and \ref{11fig:Plot for N=200 M=16} with the case of $M=8$ in Fig.~\ref{FDD fig:Plot for N=200 disp}, the max-DL gain increases from $40 \%$ to $59 \%$ to $84 \%$   as the antennas at the BS decrease from $M=16$ to $M=8$ to $M=4$. In a similar manner, the max-UL gain increases from $63\%$ to $96 \%$ to $142\%$ as the antennas at the BS decrease from $M=16$ to $M=8$ to $M=4$. The same behavior was observed in TDD as demonstrated by Figs.~\ref{TDD fig:Plot for N=200 M=4} and \ref{TDD fig:Plot for N=200 M=16} compared with Fig.~\ref{fig:TDD Plot for N=200 disp} as the max-UL gain increases from $20\%$ to $28 \%$ to $35\%$  and the max-DL gain increases from $20\%$ to $27 \%$ to $35\%$  as the antennas at the BS decrease.}}
\par
\textcolor{black}{ Zero forcing (ZF) beamforming can also be used as a heuristic in order to obtain beamforming vectors with less complexity \cite{UL_DL_duality}. Fig.~\ref{Fig: COmpare ZF} {shows the UL-DL trade-off regions} when ZF beamforming is used for FDD and TDD with $N=200$. For FDD, compared to the results of Fig.~\ref{FDD fig:Plot for N=200 disp}, the rate regions shrink when ZF is used as the maximum achievable downlink  rate is reduced from 3.26  bit/sec/Hz to 2.84 bit/sec/Hz and thus WMMSE yields an improvement of  $15\%$ over ZF. Similarly,  WMMSE yields  an improvement of  $28\%$ in terms of the maximum achievable uplink rate. The max-DL gain increases slightly from  $59\%$ under WMMSE to  $63\%$ for ZF beamforming. The max-UL gain of $96\%$ is the same under WMMSE and ZF beamforming. Hence, for ZF, the improvement in performance due to joint design compared with the uplink/downlink fixed design is similar to WMMSE. Similarly for TDD, the rate regions shrink when ZF is used as the maximum achievable downlink  rate is reduced from 3.59  bit/sec/Hz to 3.12 bit/sec/Hz and thus WMMSE yields an improvement of  $16\%$ over ZF. The maximum achievable uplink rate shows an improvement of  $28\%$ when WMMSE is adopted compared to ZF.
Furthermore, under ZF the max-UL and max-DL gains are similar to those when WMMSE is adopted for beamforming. Consequently, while ZF has lower complexity and performance compared to WMMSE, the relative performance gains of joint-design are about the same as WMMSE.}
\subsection{Complexity Analysis}
\label{complexity analysis}
\textcolor{black}{In this section, we present the computational complexity associated with the BCD algorithm of Section~\ref{sec 4}. Moreover, we compare the overhead and computational complexity associated with the optimized joint IRS design and the fixed uplink/downlink {heuristic designs}.}  \subsubsection*{Computational Complexity}
\par 
\textcolor{black}{The BCD algorithm of Section~\ref{sec 4} iteratively updates the downlink beamforming vectors, uplink combining vectors, power control  and IRS configuration blocks. We present the per-iteration complexity associated with updating the BCD blocks.}
\par 
\textcolor{black}{To update the BCD blocks, the effective downlink and uplink channel associated with user $k$ need to be computed as given in \eqref{1a} and \eqref{1b}. {For $t \in \{UL,DL\}$, computing the $k^{th}$ effective downlink/uplink channel involves $N$ operations to multiply the diagonal matrix $\boldsymbol{\Theta}_{t}$ by the reflected channel $\boldsymbol{h}_{t,r,k}$ and another $MN$ operations to multiply the aforementioned product by $\boldsymbol{G}_{t}$. Moreover,  $M(N-1)$ and $M$ addition operations are needed to compute the expression in \eqref{1a} or \eqref{1b}.  Subsequently,}  $2MN+N$ operations are needed to obtain the effective channel for each user $k$ and hence the complexity of computing the $K$ effective channels is $\mathcal{O}(KMN)$.}
\par
\textcolor{black}{ 
For given effective downlink channels $\{\boldsymbol{h}_{DL,k}\}_{k=1}^K$, obtained with complexity $\mathcal{O}(KMN)$, the downlink beamforming vectors $\{\boldsymbol{w}_k\}_{k=1}^K$ can be updated either using the WMMSE algorithm or zero-forcing (ZF). For given effective channels, each (inner) iteration of the WMMSE algorithm, presented in \eqref{WMMSE_updates}, is dominated by the matrix inversions and computing the products $\boldsymbol{h}_{DL,k}^T \boldsymbol{w}_i $ as well as the search for the optimal Lagrange multiplier $\nu$ in each iteration of WMMSE. Hence, the \emph{per BCD-iteration} complexity is $\mathcal{O}(I_{B}I_{\nu} K M^3+ I_{B}K^2 M)$, where $I_{B}$ is number of (inner) iterations needed for updating the WMMSE algorithm and $I_{\nu}$ is the number of iterations needed to search for $\nu$. Consequently, when WMMSE is adopted, the complexity of computing the effective channels and updating the downlink beamforming vectors is $\mathcal{O}(C_{DL})$ where $C_{DL} = I_{B}I_{\nu} K M^3+ I_{B}K^2 M +KMN$. Alternatively, when ZF is used to update $\{\boldsymbol{w}_k\}_{k=1}^K$, \emph{one} update is needed and is dominated by computing an $M \times M$ matrix inversion and $K$ matrix multiplications of the inverted matrix by an $M$-length vector. Consequently, when ZF is adopted, the complexity of computing the effective channels and updating the downlink beamforming vectors is $\mathcal{O}(C_{DL})$ where $C_{DL} = M^3 + KM^2 +KMN$.}
\par 
\textcolor{black}{The (same) effective uplink channels $\{\boldsymbol{h}_{UL,k}\}_{k=1}^K$, obtained with complexity $\mathcal{O}(KMN)$, are needed to update \emph{both} the uplink combining vectors $\{\boldsymbol{v}_k\}_{k=1}^K$ and power control $\{p_k\}_{k=1}^K$. For given effective uplink channels, updating the $K$ combining vectors using the MMSE receiver incurs a complexity of $\mathcal{O}(M^3 + KM^2)$. This follows since computing the $K$ combining vectors in \eqref{eq:UL_MMSE} is dominated by an $M \times M$ matrix inversion and $K$ multiplications of the inverted matrix by an $M$-length vector. Moreover, for given effective channels, the fractional programming (FP) algorithm in \eqref{chi update}, \eqref{gamma update} and \eqref{power update} is used to update the uplink power $\{p_k\}_{k=1}^K$. This incurs a complexity of $\mathcal{O}(K^2M + I_FK^2)$, where $I_F$ is the number of iterations needed to update the power control. Subsequently, the complexity of computing the effective channels and updating the uplink combining vectors  and power control is $\mathcal{O}(C_{UL})$ where $C_{UL} = M^3 + KM^2 + K^2M + I_FK^2+KMN$}. 

\par 
\textcolor{black}{To update the IRS configuration, the $\boldsymbol{\lambda}_{t,i,k}$ and ${\mu}_{t,i,k}$ expressions in \eqref{21a} and \eqref{21b} need to be computed for $t \in \{UL,DL\}$ and for $i,k \in \mathcal{K}$. This incurs a complexity of $\mathcal{O}( P K M N)$, where $P=2K$ corresponds to the $K$ (downlink) beamforming vectors and $K$ (uplink) combining vectors. When the expressions in \eqref{21a} and \eqref{21b} are given, each (inner) iteration of the RCG  is dominated by computing the $P$ Euclidean gradients in \eqref{eq: Gradients} and hence incurs a complexity of $\mathcal{O}( I_R PKN)$, where $I_R$ is the number of (inner) iterations associated with the RCG algorithm. Consequently, the complexity of computing the necessary expressions in \eqref{21a} and \eqref{21b} and updating the IRS configuration is $\mathcal{O}(C_{RCG})$ where $C_{RCG} =  I_R PK N+PKMN$.}
\subsubsection*{Convergence of the BCD algorithm}
\par
\begin{figure*}%
\centering
\begin{subfigure}{.99\columnwidth}
\captionsetup{justification=centering,margin=1cm}
\includegraphics[width=\linewidth]{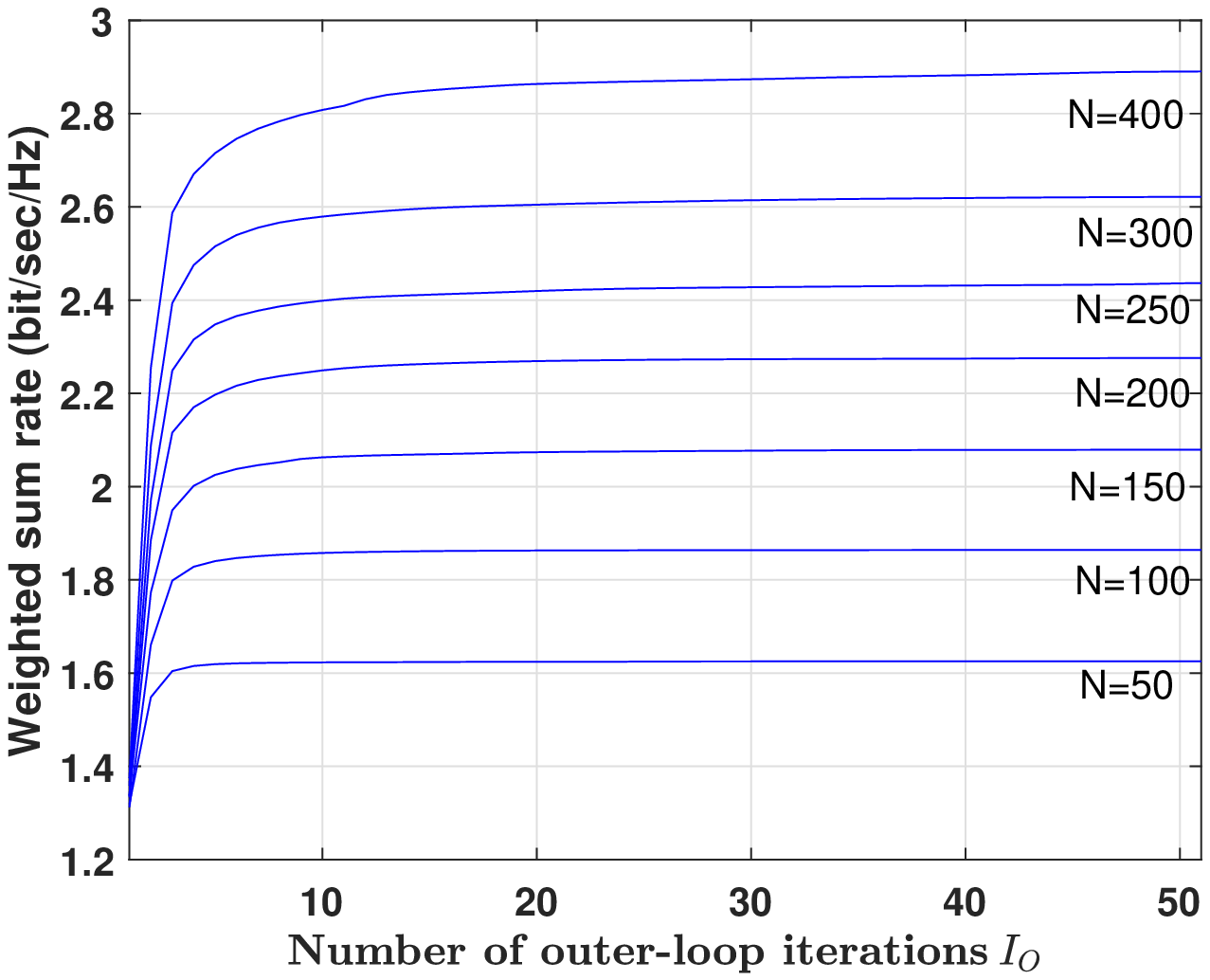}
\subcaption{Convergence for FDD system}     \label{fig:Convergenece FDD}
\end{subfigure} \hfill 
\begin{subfigure}{.99\columnwidth}
\captionsetup{justification=centering,margin=1cm}
\includegraphics[width=\linewidth]{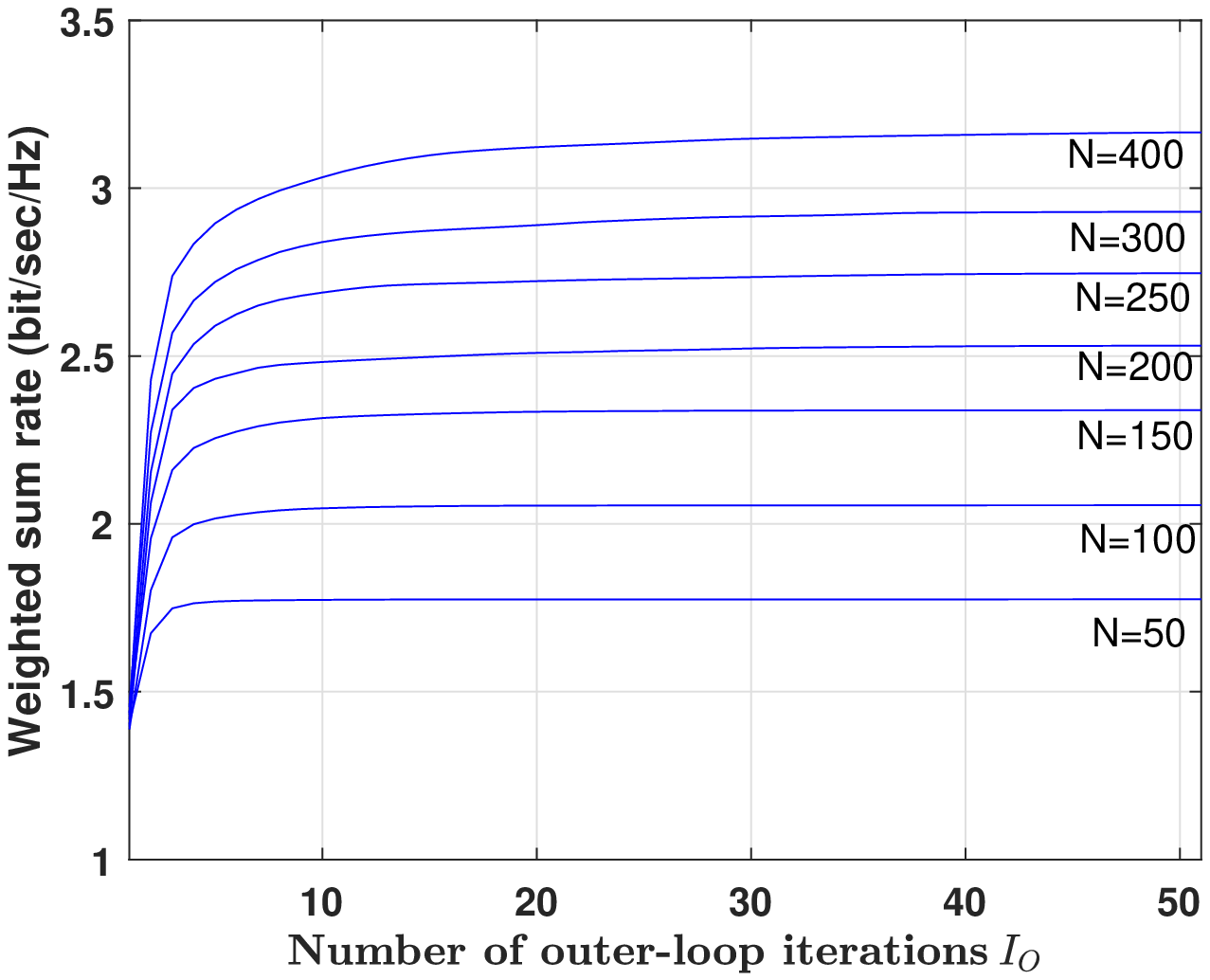}
\subcaption{Convergence for TDD system}
\label{fig:Convergenece TDD}
\end{subfigure}
 \caption{Convergence of BCD algorithm used to optimize the joint design for FDD and TDD systems.}
\end{figure*}

 
\textcolor{black}{We  consider the convergence of the BCD algorithm of Section~\ref{sec 4} for FDD and TDD systems. With $M=8$, $\alpha = \beta =\frac{1}{2}$ and hence ${\alpha\beta } ={(1-\alpha)(1-\beta)}$, Figs.~\ref{fig:Convergenece FDD} and \ref{fig:Convergenece TDD} show the weighted sum-rate $J_{WSP}$ from \eqref{eq:UL_DL_WSR} vs the number of outer-loop iterations \textcolor{black}{$I_O$} for different number of IRS elements $N$, under independent weights. It is observed that the solution converges reasonably quickly, and in all cases the solution reaches $99\%$ of the final value in less than 20 iterations. }
\subsubsection*{Complexity Comparisons}
\par
\textcolor{black}{We first compare the \textcolor{black}{three proposed designs as well as the individual design} in terms of feedback overhead and configuration periods. In FDD, the uplink and downlink transmissions occur concurrently and hence an individual design is not achievable. In TDD, an individual IRS design would constitute sending two sets of configurations to the IRS, one for downlink and one for uplink. Conversely, the optimized joint design as well as the fixed-uplink and fixed-downlink design involve sending \emph{one} IRS configuration. Thus, these designs halve the feedback overhead compared to the individual design. In practice, the individual design would also involve switching the IRS configurations between downlink and uplink and hence (compared to the optimized joint design and the fixed-uplink/fixed-downlink designs) increases power consumption and the duration of the period where the IRS system can not be used for communication.}
\par 
\textcolor{black}{From the computational complexity analysis, RCG for fixed-uplink and fixed-downlink requires only $P=K$ gradients instead of $2K$, reducing the computational complexity of this dominant term by a factor of $2$. In addition, the fixed-downlink (respectively, fixed-uplink) design iteratively optimizes the IRS configuration using RCG and WMMSE (respectively, MMSE) algorithms. Then for a fixed IRS configuration, the uplink combining vectors (respectively, downlink beamforming vectors) are computed by applying the MMSE (respectively, WMMSE) \emph{once} to the effective uplink (respectively, downlink) channels. Hence, compared to an optimized joint design, the complexity of a fixed-downlink design is reduced by $I_O-1$ iterations of complexity $\mathcal{O}(C_{UL})$, where $I_O$ is the number of outer-loop iterations and $C_{UL}$ indicates the number of operations needed to update the uplink combining vectors and power control. Similarly, compared to an optimized joint design, the complexity of a fixed-uplink design is reduced by $I_O-1$ iterations of complexity $\mathcal{O}(C_{DL})$, where $C_{DL}$ indicates the number of operations needed to update the downlink beamforming vectors.} 
\par 
\textcolor{black}{The aforementioned complexity comparisons, when combined with the trade-off regions presented in Section~\ref{Trade-off region Analysis}, demonstrate the flexibility offered by the proposed optimized joint design as well as fixed-uplink and fixed-downlink designs in trading off spectral efficiency performance for reductions in computational complexity, overhead and configuration periods. These trade-offs may be particularly appealing when the reductions in overhead and computational complexity can be achieved in exchange for marginal losses in spectral efficiency. For instance, as demonstrated in Section \ref{sec 5}-B, {the fixed-uplink and fixed-downlink designs perform almost as well as the individual design in  TDD scenarios{ under equal and proportional fair user-weighting strategies (but not independent weights).}} {Thus, under these scenarios, the fixed designs almost achieve the performance of the individual design but with lower complexity than the optimized joint design and individual design, and lower overhead/power consumption than the individual design.} }

\par 
\textcolor{black}{{On the other hand, in the FDD scenarios considered in Section V-B, the proposed jointly optimized IRS design performs substantially better than {the fixed-uplink and fixed-downlink heuristic designs} {and diminishes the gap to the individual design}. {In TDD, provided that the user weights are independent, the joint design substantially improves the spectral efficiency performance compared with the{ fixed-uplink and fixed-downlink heuristic designs.} Moreover, the relative performance improvement of the joint design over the fixed designs increases as the ratio of IRS to BS elements increases.} } Hence, for these scenarios, the complexity comparisons along with the trade-off regions allow the designer to trade-off spectral efficiency with complexity and overhead when adopting an IRS design. For example, if complexity is not a constraint, then the optimized joint design provides the highest spectral efficiency in FDD. }

 \section{Conclusion}\label{sec 6}
 \par
\textcolor{black}{This paper has investigated jointly optimized uplink-downlink IRS design for both FDD and TDD multi-user systems.{ Trade-off regions between uplink and downlink rates achieved by a jointly optimized IRS design} were obtained by formulating and optimizing a WSP. A joint design is essential to the operation of FDD systems as{ uplink and downlink transmissions occur simultaneously and hence an individual} design is not physically realizable. In TDD, a joint design can substantially reduce signalling overhead, power consumption and configuration periods.} 
\par 
\textcolor{black}{{With the system configuration optimized by the proposed BCD algorithm, the effect of different parameters on the improvement in spectral efficiency due to a jointly optimized design compared to the fixed-downlink/fixed-uplink designs was investigated. When the user-weights become less uniform and as more flexibility in resource allocation is needed, a joint design provides significant gains compared to the two heuristic designs (i.e., fixed-uplink and fixed-downlink design).}}
\par 
Further, the improvement in performance due to joint design is similar under both WMMSE and ZF beamforming strategies. \textcolor{black}{In addition, 
the optimized joint design was compared to two slicing benchmarks, slicing-with-interference 
and slicing-without-interference. This comparison revealed that the joint design significantly outperformed the two slicing benchmarks and that the interference between the slices alone was not responsible for the inferior  performance achieved by a sliced IRS.}

\par
\textcolor{black}{\textcolor{black}{ A complexity analysis of the joint and fixed designs was used along with the computed trade-off regions to gain insights into scenarios where a marginal loss in spectral efficiency can be traded off for reductions in computational complexity, configuration periods, and overhead.}  In FDD, the gains achieved by the joint design compared with the alternative schemes were significant in all considered scenarios and thus the joint design has significant benefits for FDD in all scenarios. \textcolor{black}{For TDD, the joint design yields substantial benefits compared to fixed-downlink/fixed-uplink under independent weights. Moreover, the relative performance improvement of the joint design over the fixed designs increases as the ratio of IRS to BS elements increases.} \textcolor{black}{Otherwise, in TDD, the fixed-uplink (i.e., using the optimized UL IRS design for both UL and DL transmissions) and fixed-downlink designs (i.e., using the optimized DL IRS design for both UL and DL transmissions) {achieve almost the same performance as the individual design (separately optimized IRS configurations for UL and DL) but with less complexity than the jointly optimized design, or the individual design.}}}

\ifCLASSOPTIONcaptionsoff
   \newpage
\fi

\bibliographystyle{IEEEtran}

\end{document}